\documentclass[runningheads,a4paper,oribibl]{llncs}

\usepackage[table]{xcolor}
\usepackage{epsfig}
\usepackage{amsmath}
\usepackage{amssymb}
\usepackage{booktabs}
\usepackage{stmaryrd}
\usepackage{url}
\usepackage{longtable}
\usepackage[figuresright]{rotating}
\usepackage{excludeonly}
\usepackage{bibentry}
\usepackage{subfig}
\usepackage{wrapfig}
\usepackage{rotating}
\usepackage{subfig}
\usepackage{textcomp}
\usepackage[latin1]{inputenc}
\usepackage{cspsymb}
\usepackage[english]{babel}
\usepackage{fancyvrb}
\usepackage{listings}
\usepackage{epstopdf}
\usepackage{wasysym}
\usepackage{multirow}
\usepackage{color}
\usepackage{colortbl}
\usepackage{arydshln}%for dashed vertical lines
\usepackage{eucal}
\usepackage{listings} %for verbatim reviewers' comments, but with line breaks
\renewcommand{\bf}{\textbf}

\newcommand\myseq[1]{\mathsf{Sq.}{\mytrace{#1}}}
\newcommand\sencrypt[2]{\mathsf{E}_{k_{#1}}(#2)}
\newcommand\encrypt[2]{\mathsf{E}_{pk_{#1}}(#2)} %\{#1\}_{#2(#3)}}
\newcommand\decrypt[2]{\mathsf{D}_{sk_{#1}}(#2)}

\newcommand\sign[2]{\mathsf{S}_{sk_{#1}}(#2)}

\newcommand\digballot[2]{\mathsf{DigB}(#1,#2)}
\newcommand\vote[2]{\mathsf{V}(#1,#2)}
\newcommand\raw[2]{\mathsf{Raw}(#1,#2)}
\newcommand\ballot[3]{\mathsf{B}(#1,#2,#3)} %_{sk_#1}(#2)}
\newcommand\rhs[2]{\mathsf{RHS}(#1,#2)}
\newcommand\receipt[3]{\mathsf{R}(\sign{#1}{\rhs{#2}{#3}})}
\newcommand\ind[1]{\mathsf{Ind.}{#1}}

\newcommand\mytrace[1]{\langle#1\rangle}
\newcommand\myset[1]{\{#1}
\newcommand\mysetcomp[1]{#1\}}

\newcommand\ie{i.e.\ }

\newcommand\etal{\emph{et al.}}
\newcommand\mkset[1]{\mathcal{#1}}

\newcommand{\PAV}{Pr\^et \`a Voter\ }
\newcommand{\pav}{Pr\^et \`a Voter}

\newcommand\corr[1]{{#1}}
\newcommand{\bim}[1]{\textbf{\textit{#1}}}
\newcommand{\rom}[1]{\uppercase\expandafter{\romannumeral #1\relax}}
\urldef{\mailsa}\path|muratmoran@gmail.com|
\urldef{\mailsb}\path|james@chiastic-security.co.uk|

\newcommand{\keywords}[1]{\par\addvspace\baselineskip
\noindent\keywordname\enspace\ignorespaces#1}

\begin{document}
%\label{firstpage}

\mainmatter

\title{Automated Analysis of Voting Systems under an Active Intruder Model in CSP} % Title of the article
\titlerunning{Automated Analysis of Voting Systems under an Active Intruder Model}

\author{Murat Moran\inst{1} \and James Heather\inst{2}}

\institute{William Marsh Rice University, \\
Computer Science Department\\
6100 Main St, Houston, TX 77005, USA\\
\mailsa\\
%\url{www.cs.rice.edu}
\and
Mendeley Ltd, London, UK \\
\mailsb\\
}

\date{\today}

\maketitle

% Main part of your article
% =========================

%\clearpage
\begin{abstract}This article presents a novel intruder model for automated reasoning about anonymity (vote-privacy) and secrecy properties of voting systems. We adapt the lazy spy for this purpose, as it avoids the eagerness of pre-computation of unnecessary deductions, reducing the required state space for the analysis. This powerful intruder behaves as a Dolev-Yao intruder, which not only observes a protocol run, but also interacts with the protocol participants, overhears communication channels, intercepts and spoofs any messages that he has learned or generated from any prior knowledge.

We make several important modifications in relation to existing channel types and the deductive system. For the former, we define various channel types for different threat models. For the latter, we construct a large deductive system over the space of messages transmitted in the voting system model. The model represents the first formal treatment of the vVote system, which was used in November 2014, in state elections in Victoria, Australia.
\end{abstract} % Abstract of the article

% Keywords for the article

% ---------------------------------------------------------------------------
\section{Introduction}
\label{sec:intro}
% ---------------------------------------------------------------------------

This article~\footnote{This article extends a previous version appeared in the Proceedings of the 13th International Workshop on  Automated Verification of Critical Systems (AVOCS 2013).} presents a novel intruder model for automated reasoning about anonymity and secrecy properties of voting systems. It is much stronger than the passive attacker used in previous work in~\cite{MHS14, MHS15}, as it behaves as a Dolev-Yao intruder~\cite{DY83}. This type of intruder not only observes a protocol run, but also interacts with the protocol participants, overhears communication channels, intercepts and spoofs any messages that he has learned or inferred from previous knowledge. This approach is inspired by lazy spy (perfect spy)~\cite{RG97}, which is designed for cryptographic protocol analysis, and called `lazy' as it avoids the eagerness of pre-computation of unnecessary inferences. To apply this intruder model to voting systems, several important modifications are needed in relation to existing channel types and the deductive system. For the former, we benefit from Creese~\etal~\cite{CGH+05, CGRZ03}, who defined various channels for different threat models in ubiquitous computing environments. For the latter, we construct a larger deductive system over the space of messages transmitted in the model.

The basis for our model is the vVote voting system, which is based on \pav~(`PaV') \cite{Rya04}, and developed for use in Victorian Electoral Commission (VEC) elections during early voting phase in November 2014 \cite{BCH+12a, BCH+12b, Cul13} in Victoria, Australia. In the elections, there were 40 legislative council representatives and 88 districts in 8 regions, and a mixture of the alternative vote\footnote{also called instant-runoff voting (IRV)} (AV), and the single transferable vote (STV) in order to rank these candidates. In total, 1121 votes were cast using vVote voting system. Most of the key features of PaV are retained in the vVote system. However, to adapt the system to such a complex election setup, a number of modifications have been necessary in the system design; for instance, the inclusion of distributed ballot generation, an electronic ballot marker (EBM) to assist the voter in filling out the ballot, and print-on-demand ballots for voters who are voting away from their registered polling station. 

In the literature, there has to date been no successful automated anonymity verification of voting systems using the Dolev-Yao intruder model. For example, Backes~\etal~\cite{BHM08} analysed voting systems mechanically in terms of verifiability properties. However, no automated analysis of anonymity property was provided as the ProVerif tool employed was unable to cope with algebraic equivalences, and hence, only a hand proof was given. Similarly, Delaune~\etal~\cite{DKR10, DRS08} and Smyth~\cite{Smyth2011} verified vote privacy of the FOO voting system~\cite{FOO92} with an additional compiler (ProSwapper), but these lacked a proof of its soundness---we understand this to mean that the framework may produce false negatives. However, the ProVerif verification tool is capable of evaluating reachability properties of security protocols for an unbounded number of sessions using observational equivalence. Chadha~\etal~\cite{CCK12} managed to verify the anonymity of the FOO voting system using a prototype, Active Knowledge in Security Protocols (AKISS), which was written in the OCaml programming language and implemented to check equivalences; however, the tool used was inefficient, and an important part of the analysis, the termination of the saturation procedure as required for deciding trace equivalences, was merely conjectured rather than proven. AVISPA~\cite{ABB+05} is another tool providing a support for the evaluations of reachability properties of security protocols for an unbounded number of sessions with TA4SL verification tool~\cite{BHKO04}. Lastly, Cremers~\cite{Cre08, Cre08a} proposes the Scyther tool that can evaluate reachability properties of security protocols for a bounded and unbounded number of sessions. It does not guarantee the termination of unbounded session, but it produces results for the bounded case. In our analysis, Communicating Sequential Processes (CSP) formal language and Failures-Divergence Refinement (FDR) model checker is used, which supports automated analysis of security protocols for bounded number of sessions. Although, no proof of soundness is given in this article, what we get from this tool is that it guarantees the termination for limited number of session and we do not get false negatives. In order to generalise the verification to models of arbitrary size, there are several techniques in the literature that can be employed, such as structural and data-independent induction [Ros97, Ros10]. However, such techniques do not easily apply to the models that have been developed here as the established results require rather strict conditions, which have not been satisfied in the voting system model used in this article.

This article presents a formal framework that is efficient in terms of cutting down unnecessary states and flexible for usage with privacy-related property analysis of voting systems under an active intruder model by adapting the lazy spy. Additionally, in order to demonstrate the suitability of this intruder model for evaluating voting systems, we model and analyse a real-world voting system vVote that was employed on a large scale real election held in 2014. % In this part of the investigation, the anonymity requirement is covered, as well as it being shown that a generic voting system can be analysed effectively using the lazy spy in CSP with the FDR model checker~\cite{GGH+}.

The rest of the article is structured as follows. Section~\ref{sec:vvote} presents an overview of the vVote voting system. In Section~\ref{sec:model}, the vVote voting system is modelled in CSP and the lazy spy intruder model is further extended for the analysis of voting systems. Section~\ref{sec:analysis} analyses the vVote voting system model regarding the formal specification of anonymity given in~\cite{MHS14}, investigates the analysis of the model under alternative assumptions, such as the presence of a corrupt election authority and then proposes applicability of the framework in secrecy analysis of voting systems automatically using FDR as the model checker. Section~\ref{sec:conclusion} presents conclusions and discusses the findings.
% ---------------------------------------------------------------------------
\section{vVote System Outline}
\label{sec:vvote}
% ---------------------------------------------------------------------------
Over the last few decades many trustworthy voting systems have been proposed. However, only a few have been deployed in large-scale real elections. Regarding these, Scantegrity~\rom{2}~\cite{CCC+08} was the first E2E voting system deployed in a binding governmental election on November 3, 2009~\cite{CCC+10}, involving a relatively small electorate with 1728 voters. Additionally, Norway used an internet voting protocol~\cite{Gj10} in the municipal elections in September 2011, in which more than 25,000 voters cast their votes using this protocol. Moreover, STAR-Vote~\cite{BBK+12} is another voter verifiable DRE-style voting system, which is going to be used in Travis County, Texas, the United States, involving over 450,000 registered voters. 

The vVote voting system is an  end-to-end (E2E) paper-based electronic voting system based on \pav~\cite{CRS05,Rya04}. However, a number of modifications have been made to the original \PAV system. The main difference is that an electronic device is deployed in order to facilitate accommodating a candidate list with over 30 candidates on the ballot forms. This also helps voters to indicate their preferences among many other candidates. However, deploying these computers requires further trust on them as they know about the voters' choice at the time of voting. Hence, a misbehaving device can violate voter anonymity or vote privacy and as a result, for further confidence in the design of this promising real-world voting system, formal verification is needed.

\begin{figure} %[htp]
  \centering
    \includegraphics[scale=1.1]{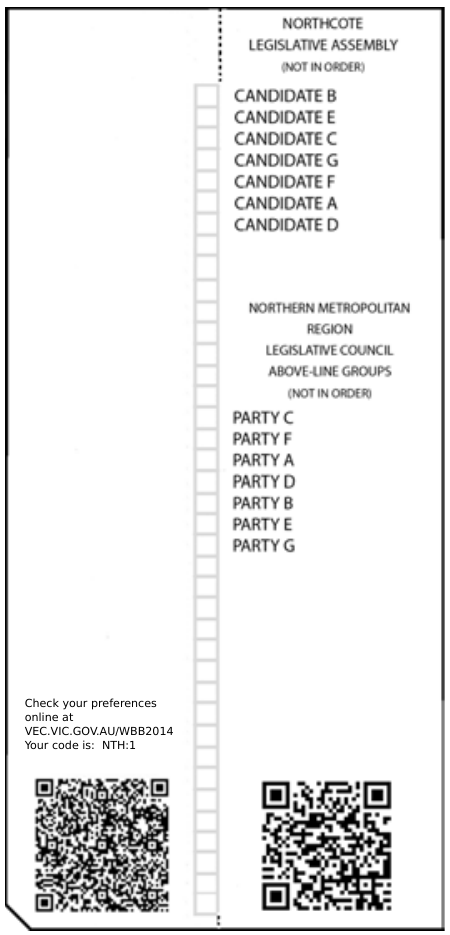}
   \caption{vVote ballot form}
   \label{fig:vvoteballot}
\end{figure}

Figure~\ref{fig:vvoteballot} illustrates the vVote ballot form. On one side there is a randomly permuted candidate list and a QR code at the bottom that records the permuted candidate order, on the other are marking boxes, a unique serial number and another QR code, corresponding to the onion that embeds the candidate order. vVote gives the voters ability to verify if their vote has been included in the final tally with a printed record of their vote. In the middle of the ballot, there is a perforation line. Once the voter marks her choices in the vote boxes, she then tears down the perforation line, keeps the half with a copy of the marked boxes as receipt, signed by the system, and shreds the other half of her ballot with the candidate order. As the candidate order is randomised and shredded, the receipt with the mark on does not reveal how the voter voted. The cast ballots are then made published on the Web Bulletin Board (WBB), allowing voters to check whether their vote has been included in the final tally.

%\begin{wrapfigure}{r}{0.5\textwidth} %{l} for left allign
%\end{wrapfigure}

%In the construction of onions, exponential and normal ElGamal public key algorithms are used. 
%For instance, for the legislative council election, a value in $G_q$ is chosen to represent each party name. Suppose the value is $\alpha$, then the corresponding onion for this party is the ElGamal term $\encrypt{}{\alpha} = (g^r, \alpha h^r)$, where $h = g^x$, $r \in \mathbb{Z}^{\ast}_{q}$ is a random value and $x \in \mathbb{Z}^{\ast}_{q}$ is a random secret value.

\begin{figure}%[htp]
  \centering
    \includegraphics[scale=0.55]{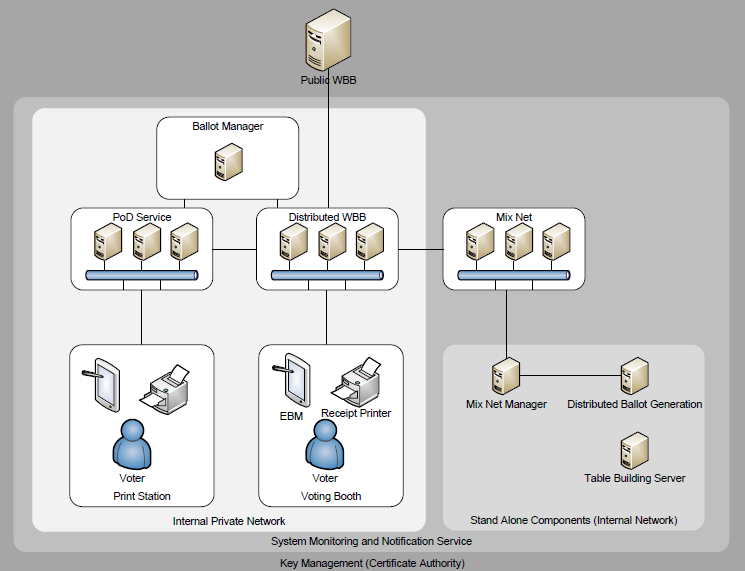}
   \caption{vVote design overview}
   \label{fig:vvotedesign}
\end{figure}

Figure~\ref{fig:vvotedesign} depicts the overview of vVote design including the main components, some of which will be covered in the following paragraphs, such as, Print on Demand (POD) Service, Distributed Ballot Generation and Mixnet Manager.

%\begin{wrapfigure}{r}{0.5\textwidth} %{l} for left allign
%\end{wrapfigure}

\subsection{Voting Ceremony}
Having registered with the poll worker, the voter or the former interacts with the system to get a ballot paper. A printer will be deployed that derive the permuted candidate list on the ballot form when it is actually being printed in the polling station. On the printed vVote ballot form, there is a QR barcode that consists of a digitally signed serial number. The association between the candidate order and the serial number on the ballot should be kept secret. The voter now can choose to audit her ballot form by checking the encrypted list of candidates on the WBB matches the plaintext candidate order on the ballot form. Once she is ready to vote, she scans the barcode on the ballot to the EBM (a new front-end component where the voter can see and cast her ballot form in an electronic environment). Once the voter completes marking on the EBM, she gets her printed marked ballot form from the EBM. The voter now should separate the candidate list from the ballot form and dispose it in order to ensure receipt-freeness. The EBM interacts with the WBB to submit the vote and receive a digitally signed receipt for it, which is then printed for the voter for verification purposes by a receipt printer in the booth. The EBM submits voter preferences and the QR code to the WBB. The WBB commits, records and broadcasts the ballot data generated during the election and also signs the serial numbers allocated by the ballot manager, thereby ensuring their uniqueness. Once the voter has cast her vote and received her receipt signed by the WBB, she then leaves the polling station.

In the following subsections the election phases covering the vVote system components and protocols are explained.

\subsection{\bf{Pre-election}}
The pre-election phase covers the preparation of election material before the polling station opens. In this period, digital ballots are generated in a distributed fashion that are encrypted under the print-on-demand (POD) service's and the election authority's public key, before being committed to the public bulletin board. Each component like the EBM and print stations requires a public key pair in order to sign data. To allocate such public key certificates, an internal Certificate Authority (CA) is deployed as the system is offline.

Additionally, mixnets are set up by exchanging configuration files between each mix server. This ensures that each mix server is aware of one another and able to communicate each other. Key generation is also performed in this phase. All fixed values, such as; digital ballots, candidate identifiers, race identifiers, should be committed to the WBB before the election.

\subsubsection{Distributed Ballot Generation}
Ballot generation can be realised on the machine that prints the ballot form in the booth or in a distributed fashion that is similar to that described in~\cite{RT10,Rya06}, \ie a number of candidate list mixers shuffle the encrypted candidate names for each vote, which ensures that the candidate ordering is random and not generated by a single party. This eliminates single point failures. In the distributed version, a list of encrypted ballots, including a serial number, the onion encoding the candidate list, and the list of encrypted candidate names for the printer with a proof of correspondence is produced. The onion encoding the candidate list are generated by a set of candidate list key sharers, called the POD service in such a way that the candidate ordering is random by each contributing to the cryptographic values kept encrypted throughout. The key shares of the tabulation/decryption tellers responsible for the final decryption are known to the POD service and used in the construction of the encrypted ballot forms. As a result, all the candidate list key sharers have to collude in order to determine the seed values. In order for a printer (POD client) to obtain the candidate list, it generates a blinding factor, encrypts it under the POD service public key, and sends it to the POD service with a proof of knowledge. Afterwards, having received the encrypted candidate list blinded by itself, the printer removes the blinding factor, and prints the candidate list.

\subsection{\bf{During the Election}}
This phase starts with polling stations opening and lasts until the election is closed, with no further votes being allowed to be cast. The POD service allocates and transfers pre-prepared digital ballots to the WBB, and during the election the print station in the polling booth retrieves the candidate permutation from the WBB for the assigned QR code. The voter is then able to print out her ballot form. However, to ease the marking for the elections with so many candidates, the ballot data can be transferred to an electronic environment. In this new system, the electronic ballot marker (EBM) is particularly interesting as it forms the key distinctive characteristic of vVote, being the only device in the system that knows how a particular voter has voted. That is, when the voter transfers her actual ballot form to the EBM, the candidate list on her form is also transferred to the EBM, while it is destroyed and kept secret in \pav. One of the assumptions made in~\cite{BCH+12a,BCH+12b,Cul13} is that the POD client, where physical ballot form is printed, and the EBM, are located in a private environment, such as, a voting booth. During the election, the EBM interacts with the WBB to submit the vote and receive digitally signed receipt. This receipt is printed out for the voter by the POD client in a private environment for the verification of her vote's inclusion in the final tally. We now describe the POD service and protocol in detail. 

% ---------------------------------------------------------------------------
\subsubsection{vVote POD Service and Protocol}
\label{podprotocol}
% ---------------------------------------------------------------------------
The POD service (also called candidate list key sharers) provides distribution of digital ballots in a distributed manner to the polling stations in any district. As the digital ballots are prepared and committed to the WBB before the election, this service facilitates the print-on-demand ballot distribution in real time (The details about the POD service and any other part of the vVote system can be found in the software design technical report~\cite{Cul13}. Despite the fact that it is still being updated in respect of design changes, it is considered to be a natural stable description for use in our analysis).

In the ballot generation procedure, the randomised candidate order of a ballot is encrypted under the election public key $pk_{EA}$ and it is then transformed to an encryption under the POD service's public key $pk_{PS}$ without revealing the underlying message, as described in~\cite{Jak99}. The same transformation technique is also used in the POD protocol to transform the encryptions on the digital ballots into the designated POD client's public key $pk_{PC}$ and these transformed ciphertexts cannot be decrypted by anyone other than the designated printer. Other entities covered in the POD protocol are as follows:

\paragraph{Poll Worker:} a combined abstraction of the poll worker and the system that the poll worker is using.

\paragraph{POD Client:} an online printer to print out voter receipt, and located in a private environment.

\paragraph{Ballot Manager:} a server whose only role is to apportion serial numbers. It does not need to be trusted, as its assignments are backed by the WBB.

In more detail, when the voter authenticates in the polling station, the poll worker requests a random nonce-like $sessionID$ from the administrator machine. Following this, the poll worker sends the $sessionID$ to the POD service, where it is signed and stored. Having received the signed $sessionID$, the poll worker hands it to the voter in barcode form. Then, the voter scans the barcode to the POD client, which signs the $sessionID$ and submits it to the POD service. Subsequently, the POD service signs the district and the $sessionID$ and submits them to the ballot manager. Following that, the ballot manager finds the next available serial number for that district, assigns it to the submitted $sessionID$, and notifies the WBB for this assignment by signing them using its secret key $sk_{BM}$. The WBB then sends a confirmation of this assignment to the ballot manager, which returns this to the POD service. Now, the transformation of the public keys that encrypt the ballot form takes place from $pk_{PS}$ to $pk_{PC}$. Finally, the POD service signs the serial number and sends along with the transformed ciphertexts to the POD client, which can then decrypt these and print the actual ballot for the voter. 

%
%\begin{figure}[htp]
%\centering
    %\includegraphics[scale=0.7]{PoDChart.pdf}
    %\caption{Message sequence chart of POD protocol}
    %\label{fig:podprotocol}
%\end{figure}
%

\subsection{\bf{Post Election}}
Post election is the phase where the cast votes are mixed by the mixnets, decrypted and tallied by a set of key sharers, such that only a threshold set of these sharers can perform decryption.

\subsubsection{Mixing}
\label{subsub:mixing}
After all votes are received by the WBB, each mix server should be able to import the cast votes from the WBB and check the digital signatures posted to the WBB at the end of each day. As the vote data submitted from the EBM to the WBB is raw, the data is combined with the ciphertexts that were submitted to the WBB during the ballot generation by matching the serial numbers. In order to reduce the number of mixes Vote Packing is used. That is, all the votes will be mixed together even if they are from different districts. All mixes and decryption are provided by the Mixnet manager. It also provides feedbacks of possible errors or connectivity problems to the end user. Following this, the decrypted values are looked up in the vote packing table created previously. Once the Mixnet has finished mixing and decrypting, the plaintext votes with relevant proofs of shuffles are committed to the WBB (the details of the cryptographic primitives that are used by the Mixnet, such as, zero-knowledge proofs and homomorphic encryptions are abstracted away in our analysis as we assume that the ZKPs can not be faked).

\subsubsection{Tallying}
In the protocol, exponential ElGamal public key algorithm~\cite{ElG84} is used to encrypt plaintext votes. Thus, the triple $(p,g,g^x)$ will form the public-key, and $(p,g,x)$ the secret key of an agent. The subsequent encryption of a message, $m$, can be calculated as $(g^r, m h^r)$, where $h = g^x$, $r \in \mathbb{Z}^{\ast}_{q}$ is a random value, and $p$ and $q$ are large random prime numbers, such that $p = 2q+1$ holds. In the ballot form, there is a ciphertext next to each candidate. In order to decrypt the ciphertext and to obtain $m$, one should compute $(g^r)^{-x}{m h^r}$ using the secret value $x$ as $(g^r)^{-x}{m (g^x)^r} = m$. Three different tallying methods for vVote are provided in~\cite{BCH+12a}. However, in our modelling a generic decryption teller is modelled for simplicity. It decrypts and counts the votes for each candidate encrypted under the election authority's public key, and the final results for each candidate are announced by this teller once there is no more vote to count.

%In the next section the vVote voting system is modelled in CSP as well as there being description of the adaptation of the lazy spy intruder model for the analysis of voting systems. 

% this file is called up by FAOC_revised.tex
% content in this file will be fed into the main document
% ---------------------------------------------------------------------------
\section{Communicating Sequential Processes}
\label{sec:csp}
% ---------------------------------------------------------------------------
CSP is a formal language, designed to describe concurrent systems in terms of components that interact by means of message passing. CSP allows us to model systems in terms of \emph{processes}, which can synchronize and interact with the environment. Besides, it provides several semantic models to analyse the behaviour of processes and systems.

In CSP, processes are defined in terms of \emph{events} that the process can perform. A synchronised event can happen when all processes agree on executing it; it happens when it is inevitable. The set of events that are visible is called $\Sigma$. Processes are associated with an interface or {\em alphabet}, denoted $\alpha P$. If no alphabet is explicitly defined then it will be the set of events that the process can perform. The simplest process is $\STOP$, which fundamentally does nothing. $\SKIP$ is another named process, which terminates immediately. However, it is not a deadlock as in $\STOP$, but a successful termination.

Given a process $P$ and an event $a$ in $\Sigma$, the prefix process $a \then P$ is initially willing to perform an event $a$. Therefore, it waits until the event, $a$, is performed then behaves like the process $P$. Events can also be structured into any number of parts.  For example, an event of the form $c.v$ can represent a channel $c$ passing value $v$. The set of values $T$ that can pass along $c$ is the {\em type} of $c$, so the set of events associated with channel $c$ of type $T$ is $\{ c.v \mid v \in T\}$.  This is also written $\{| c |\}$.  If $C$ is a set of channels, then $\{| C |\} = \Union_{c \in C} \{| c |\}$. The input process $c?x \then P(x)$ is initially prepared to accept a value that will be bound to the locally introduced variable $x$ along channel $c$, and then behave as $P$ having received input $x$.  

CSP offers choice operations for processes, which are called \emph{external} and \emph{nondeterministic} choice operators denoted as $\extchoice$ and $\intchoice$ respectively. The process $P \extchoice Q$ can act like $P$ or $Q$ depending on the choice of the initial event chosen by the environment. For instance, for the process $(a \then P) \extchoice (b \then Q)$, if the first event chosen is $a$ then the process will behave as the process $P$, after performing the event $a$. Similarly, if the first event chosen is the event $b$, subsequently the process will act as the process $Q$. While the external choice operator leaves the choice to its environment, in a nondeterministic process, the choice is made internally. Thus, the process $(a \then P) \intchoice (b \then Q)$ can act as either $a \then P$ or $b \then Q$ and the environment has no control over which. Indexed versions of external and nondeterministic choices
allow the choices to be made among a number of processes.

Systems can be made up of a collection of processes that run in parallel and synchronise on the events that they agree to perform. Alphabetised parallel $P \parallel Q$ executes $P$ and $Q$ in parallel, where they have to synchronise on those events that are in both of their alphabets, but they can perform other events independently.
Thus, they must only agree on the events in the intersection $\alpha P \inter \alpha Q$.   This operator is associative and commutative, so we can combine any number of processes in parallel in any order without ambiguity.  Thus we may write $P \parallel Q \parallel R$ for the parallel combination of three processes. Alternatively, we may wish to run any two processes independently of each other, i.e., they do not synchronise on any events, not even those that they share. The interleaving operator is written ``$\interleave$''.  This also has an indexed form to describe the interleaving of a family of processes (further details on CSP can be found in~\cite{RSG+00, Ros10}).

% ---------------------------------------------------------------------------
\section{vVote and Intruder Model}
\label{sec:model}
% ---------------------------------------------------------------------------
%% ---------------------------------------------------------------------------
%\section{Modelling vVote and Active Intruder}
 %\label{sec:dy:modelling}
%% ---------------------------------------------------------------------------
Conventionally, security protocols consist of several agents sending messages to each other on the medium they share or on direct communication channels. The vVote voting system is modelled in terms of a number of agent processes that run in parallel and these processes behave as the corresponding components of the voting system~\footnote{CSP code for vVote model and the adaptation of lazy spy into the analysis from which the experimental results given in this article can downloaded from the principal author's personal webpage http://muratmoran.wordpress.com/publications/ under the CSP codes title}.

In the following subsections, the messages sent on the channels of the model are defined. Secondly, the different kinds of channels that are needed for the analysis are introduced along with the process definitions for each agent, which compose the voting system model. Following this, the lazy spy intruder model acting as a Dolev-Yao intruder is adapted to analyse such voting systems. Finally, the system model and active intruder model are put together in order to reason about the system as a whole later in the analysis section.

% ---------------------------------------------------------------------------
\subsection{Data-types and Messages}
\label{functions}
% ---------------------------------------------------------------------------
Cryptographic primitives, such as encryptions and signatures, are modelled as symbolic objects like the agents, the public and secret keys, the nonces and serial numbers. For instance, encryption: $\encrypt{}{f}$, decryption: $\decrypt{}{f}$, signature: $\sign{}{f}$. Additionally, apart from these, the other messages, which can be a collection of these cryptographic primitives, are also modelled as the data-types. In this respect, the message including a serial number and an encrypted candidate list (called raw ballots here) is denoted as $\raw{s}{\encrypt{}{l}}$, and a digital ballot message formed by a signed serial number and an encrypted candidate list is modelled as $\digballot{\sign{}{s}}{\encrypt{}{l}}$. Similarly, a ballot form consisting of a candidate list, a serial number, and an index value, is $\ballot{l}{s}{\ind{i}}$; a message with a serial number and an index value forming the marking boxes on the ballot form, called $\emph{castrhs}$, which is demonstrated as $\rhs{s}{\ind{i}}$; and a receipt is the signed $\emph{castrhs}$ denoted as $\receipt{}{s}{\ind{i}}$. Finally, a message consisting of an index value and an encrypted candidate list is called a $vote$ and shown as $\vote{\ind{i}}{\encrypt{}{l}}$. Figure~\ref{fig:messages} depicts how these messages are composed in the model.     

In order to compose these messages, the model consists of several finite sets of facts, $\mathcal{F}$, as listed below. The abbreviation $W$ stands for the web bulletin board, $T$ is Tom, the poll worker, $EA$ is the election authority, $PS$ and $PC$ are the POD service and client, respectively, and $BM$ is the ballot manager. For convenience, names are abbreviated as follows: the set of candidates as $\mathcal{C}$, voters as $\mathcal{V}$, agents as $\mathcal{A}$, serial numbers as $\mathcal{S}$, nonces as $\mathcal{N}$, and public-keys and secret-keys as $\mathcal{PK}$ and $\mathcal{SK}$, respectively.
%
%\begin{figure}[htp]
%\begin{center}
\begin{tabbing}
$\mathcal{C} \quad $\=$= \{Archimedes, Babbage\}, \mathcal{V} = \{ Alice, Bob, James\}$\\
$\mathcal{A} $\>$ = \Union(\mathcal{V}, \{ $\=$Tom, authority, wbb, teller, podservice, $\\
\>\>$ podclient, ballotmngr, ebm, printer\})$\\
$\mathcal{S}         $\>$ = \{ s_1, s_2, s_3\}, \mathcal{N} = \{ n_a, n_b, n_c\}$\\
$\mathcal{PK}     $\>$ =  \set{pk_A \,\mid\, A \in \set{W, T, EA, PS, PC, BM} }$\\ %\{ pk_W, pk_T, pk_{PS}, pk_{PC}, pk_{BM}, pk_{EA}\}$\\
$\mathcal{SK}     $\>$ =  \set{sk_A \,\mid\, A \in \set{W, T, EA, PS, PC, BM} }$%\{ sk_W, sk_T, sk_{PS}, sk_{PC}, sk_{BM}, sk_{EA}\}$
\end{tabbing}
%\caption{Sets of facts}
%\label{fig:sets}
%\end{center}
%\end{figure}
The agents send various kinds of messages to each other, which need to be defined in terms of data-types. The messages mentioned above and illustrated in Figure~\ref{fig:messages} form the message set $\mathcal{M}$. The names of the sets are indicative of what messages they represent. However, to remove the ambiguity; $castrhs$ represents the cast ballots, $\mathcal{L}$ is the set of all possible candidate lists, and $\mathcal{I}$ is the set of indices with how the voter is modelled to fill in the marking boxes for her preferred candidate. 

\begin{figure}%[htp]
\centering
\begin{tabbing}
$signednonces \quad$ \=$= \myset{\sign{}{n}}\mysetcomp{$\quad\quad\quad\quad\quad\quad\quad\quad\quad\=$|\,$\=$ sk \quad$\=$\,\leftarrow $\=$\mathcal{SK}, $\\\>\>\>$n $\>$\leftarrow $\>$\mathcal{N}}$
\\
$signednonsers $\>$= \myset{\sign{}{s,n}} \mysetcomp{$\>$|$\>$sk $\>$\leftarrow $\>$\mathcal{SK}, $\\\>\>\>$ s $\>$\leftarrow $\>$\mathcal{S}, $\\\>\>\>$n $\>$\leftarrow $\>$\mathcal{N} }$
\\
$rawballots $\>$= \myset{\raw{s}{\encrypt{}{l}}}\mysetcomp{$\>$|$\>$ s $\>$\leftarrow $\>$\mathcal{S}, $\\\>\>\>$ pk $\>$\leftarrow $\>$\mathcal{PK}, $\\\>\>\>$l $\>$\leftarrow $\>$\mathcal{L}}$
\\

$digitalballots $\>$= \myset{\digballot{\sign{}{s}}{\encrypt{}{l}}}\mysetcomp{$\>$|$\>$sk $\>$\leftarrow $\>$\mathcal{SK}, $\\\>\>\>$ pk $\>$\leftarrow $\>$\mathcal{PK}$\\\>\>\>$ s $\>$\leftarrow $\>$\mathcal{S}, $\\\>\>\>$l $\>$\leftarrow $\>$\mathcal{L}}$
\\
$indices $\>$=\myset{\ind{i}} $\>$|$\>$ i $\>$\leftarrow $\>$Int \}$\\
$\it{ballotforms} $\>$= \myset{\ballot{l}{s}{\ind{i}}}$\>$|$\>$l $\>$\leftarrow $\>$\mathcal{L},$\\
                                                          \>\>\>$ s $\>$\leftarrow $\>$\mathcal{S},$\\
                                                          \>\>\>$ \ind{i} $\>$\leftarrow$\>$ \mathcal{I},$\\
                                                          \>\>\>$ a $\>$\leftarrow $\>$\mathcal{A}\}$
\\
$castrhs $\>$= \myset{\rhs{s}{\ind{i}}} $\>$|$\>$s $\>$\leftarrow $\>$\mathcal{S},$\\
                                                          \>\>\>$ \ind{i} $\>$\leftarrow $\>$\mathcal{I}\}$
\\
$receipts $\>$=\myset{\receipt{}{s}{\ind{i}}} $\>$|$\>$s $\>$\leftarrow $\>$\mathcal{S},$\\
                                                          \>\>\>$ \ind{i} $\>$\leftarrow $\>$\mathcal{I},$\\
                                                          \>\>\>$ sk $\>$\leftarrow $\>$\mathcal{SK}\}$
\\
$votes $\>$=\myset{\vote{\ind{i}}{\encrypt{}{l}}} $\>$|$\>$\ind{i} $\>$\leftarrow $\>$\mathcal{I}, $\\
                                                                        \>\>\> $pk $\>$\leftarrow$\>$ \mathcal{PK}, $\\
                                                                        \>\>\> $l $\>$\leftarrow$\>$ \mathcal{L}\}$
\\
$\it{atomicfacts} $\>$=\myset{f} $\>$|$\>$ f $\>$\leftarrow $\>$\union\{\mathcal{V}, \mathcal{N},  \mathcal{I}\}\}$
%\\
%$messages $\>$=\Union_{i \in 1..9} message_i$
%\\
\end{tabbing}
\caption{Message types used in the modelling}
\label{fig:messages}
\end{figure}
%

%In the next section, the communication channels, on which these messages are transmitted, are described. 
% ---------------------------------------------------------------------------
\subsection{Channel Types}
\label{channels}
% ---------------------------------------------------------------------------
The channels have the form $\mathcal{A}.\mathcal{A}.\mathcal{M}$, where $\mathcal{A}$ is the set of agents and $\mathcal{M}$ is the set of messages that agents may wish to transmit over the channels and these are listed in Figure~\ref{fig:messages}. 

The original framework introduced in~\cite{Low95} analysing the Needham Schroe-der Public Key (NSPK) protocol involves only (InS) \emph{Insecure channels}, \ie the whole network is not secure, and hence, any message can be manipulated in many ways by the intruder. The intruder can block, overhear and spoof any message transmitted on the insecure communication channels between the legitimate agents. This kind of communication channels are directly connected to the intruder using the renaming operator in CSP. Hence, there is no restriction in the intruder process about what he can or cannot perform on the insecure communication channels. In order words, he can act as the Dolev-Yao intruder model on such channels.
%
%\begin{figure}[htp]
%\centering
    %\includegraphics[width=\columnwidth]{Figures/inschannel}
    %\caption{(InS) Insecure Channel}
    %\label{fig:inschannel}
%\end{figure}
%

Such an assumption is too strong for voting systems that require an environment for the voters to be able to vote privately, such as a voting booth, at least if the action of receiving a ballot form is modelled as a message. This is also the case for most of the remote voting systems, where it is assumed that no one is watching over the voters' shoulder while she is casting her vote. Hence, this necessitates the existence of private channels in the voting system model. To this end, the agents in the model are enabled to communicate over a \emph{secure channel} (S), called $scomm$, on which the intruder has no power at all. \corr{For instance, when the voter is given the ballot form by the poll worker, messages including the sensitive data regarding the candidate order, are transmitted over $scomm$ channels. In the modelling of such channels, different channel names, like $scomm$, are used to distinguish the secure channels from others in order to hide the crucial information from the intruder. As stated previously, the intruder's ability is modelled using the renaming operator in the process definition of the intruder. This is to say that the intruder can perform all his bad behaviour on the channels that are connected to his process definition using renaming. Hence, the secure channel $scomm$ is shared only between honest agents, and not with the intruder. 

We here assume that at least two eligible honest voters are able to vote, and the cast votes are tallied at the end of the election. That is because an attack regarding the voter's privacy occurs in which the intruder blocks all the communication channels except the one on which the target voter communicates in order to cast her vote. Thus, the intruder would learn how the voter has voted. Therefore, at least two honest voters should be able to cast their votes without any blocking so that the intruder cannot deduce how each of them has voted (it will be explained further in Section~\ref{sec:conclusion}). This assumption requires that there exists a channel in the voting system model such that the communications made by these two honest voters with the other agents are \emph{No Spoofing and Blocking} (NSB) channels modelled as $nsbcomm$ here, and they are combinations of two different channel types; No Blocking (NB) and No Spoofing (NS) channels. On such channels the intruder can overhear the communication, but cannot block its occurrence and spoof any messages. Creese~\etal~\cite{CGH+05} describe various kinds of channels for pervasive computing environments. For instance, the \emph{No OverHearing} channel $c$ ($\it{NOH_c}$) is that which cannot be overheard, the \emph{No Blocking} channel $c$ ($\it{NB_c}$) is the channel that cannot be blocked and the \emph{No Spoofing} channel $c$ ($\it{NS_c}$) is the channel type that cannot be spoofed. The three $\it{NOH_c}$, $\it{NB_c}$ and $\it{NS_c}$ form the secure channels $scomm$ in the modelling. The NB channels in CSP are modelled when the intruder process is renamed to take/block messages from the channels on the network.
%
%\begin{figure}[htp]
%\centering
    %\includegraphics[width=\columnwidth]{Figures/nbchannel}
    %\caption{(NB) No Blocking Channel}
    %\label{fig:nbchannel}
%\end{figure}
%

Using CSP the set of messages that make sense to the protocol (they are from real communications between agents), called $comms$, can be defined as the union of sets of data objects for each message type. For instance, the following defines the vote messages sent by one agent to another. 
\begin{tabbing}
$\it{commVotes} = \{a.b.m \mid m \leftarrow votes, a\leftarrow \mathcal{A}, b\leftarrow \mathcal{A}, a \neq b\}$
\end{tabbing}
These are also useful when the intruder is afforded the ability to modify the messages on the insecure channels or not to block and fake certain data from specific agents as it may be confusing as to whether the message is already known or has just been learned from the real communication that the intruder overhears. This is used in modelling the intruder by defining the set of legitimate insecure messages sent from one agent to another $\it{Ucomms}$. Similarly, insecure NB messages $\it{Nbcomms}$ from real communication can be defined so and later used to determine what the intruder can overhear, spoof but not block.

\corr{Although, the existence of NB channels solves one problem, which is the unwanted privacy attack previously mentioned, there is another plausible attack where the intruder does not block the messages on NB channels, but can later modify and spoof the messages, \ie the intruder cannot take/block, but he can still fake messages overheard from the NB channels. Hence, if the intruder can modify and spoof one of the messages sent from one of those honest two voters, he can then deduce the other private message by looking at the election result as in the previous attack. Therefore, there is a need for a channel that cannot be spoofed, called \emph{No Spoofing} (NS) channels. On NS channels, the intruder can overhear but cannot block or spoof messages. This is exactly what we need in order to allow two honest voters to cast their votes without any interruption and modification. As such a channel is no blocking and spoofing channel, it will be called as \emph{No spoofing and blocking} (NSB) channel from now on, and in the CSP definitions of the voting system and intruder model it will be expressed as $nsbcomm$.}
%
%\begin{figure}[htp]
%\centering
    %\includegraphics[width=\columnwidth]{Figures/nschannel}
    %\caption{(NS) No Spoofing Channel}
    %\label{fig:nschannel}
%\end{figure}
%

\corr{As mentioned earlier, secure channels are combinations of NB, NS and NOH channels. NOH channels are the channels that the intruder cannot overhear any messages on. On such channels the intruder can block and spoof messages, but cannot overhear the communication channel. The implementation of this channel in CSP is similar to the others'---it is modelled by restricting what the intruder can overhear with a defined set of network messages.}
%
%\begin{figure}[htp]
%\centering
    %\includegraphics[width=\columnwidth]{Figures/nohchannel}
    %\caption{(NOH) No OverHearing Channel}
    %\label{fig:nohchannel}
%\end{figure}
%

\corr{For the analysis of vVote later in this article, we need to define what information flows over: secure channels $scomm$, insecure channels $comm$, and no spoofing and blocking channels $nsbcomm$, because of the reasons explained previously. This can be done in two ways: the first one is that all agents on the network work on insecure communication channels ($comm$) in which case secure, and no spoofing and blocking channels need to be defined. The second way is that all agents communicate over a NSB channel ($nsbcomm$), and the secure and insecure channels are defined accordingly. As we know what information should be shared with the intruder, it is easier to define the insecure communications than defining the others, meaning that the second way of defining channels is the one to follow for the ease of modelling. This will also help to reduce the size of the required state space for automated analysis. In terms of the deduction system that is used in the model, following the second way does not have any impact on the deductions that may be made by the intruder because the same set of information is given to the intruder and the deduction system remains the same in each case. Table~\ref{tab:channeltypes} illustrates the intruder's capabilities on different channels used in this analysis.}

\setlength{\extrarowheight}{.75ex}
\begin{table}%[htp] %{}
%\small
\centering
\begin{tabular}{>{\columncolor[gray]{.8}[5pt]}lccccc}
\rowcolor{gray!50}  &  \multicolumn{1}{c}{Secure} & \multicolumn{1}{c}{No OverHearing} & \multicolumn{1}{c}{No Spoofing and Blocking}  &  \multicolumn{1}{c}{Insecure }\\\noalign{\vskip -0.5pt}
\rowcolor{gray!50}  &  \multicolumn{1}{c}{(S)} & \multicolumn{1}{c}{(NOH)} & \multicolumn{1}{c}{(NSB)}  & \multicolumn{1}{c}{(InS)} \\\noalign{\vskip 1pt}
\rowcolor{gray!15} overhear & X & X & $\tick $ &  $\tick $ \\\noalign{\vskip 1pt}
\rowcolor{gray!15} block & X &$\tick $ & X & $\tick $ \\\noalign{\vskip 1pt}
\rowcolor{gray!15} spoof & X &$\tick $ & X & $\tick $ \\\noalign{\vskip 1pt}

\end{tabular}
\caption{The intruder's capabilities on different channels}
\label{tab:channeltypes}
\end{table}}
 
%It is not expressed as a proper channel type here, but instead is modelled as a special power given to the intruder in his CSP process definition, where he can overhear, block and spoof messages --- it could equally be modelled such that all channels are insecure, and the NSB and private channels are the subset of these insecure channels, whereby the intruder would be restricted to define what he is capable of doing as in~\cite{CGH+05}. However, in this modelling, it is more convenient, first, to model the private and NSB channels, restricting the intruder and then give him more power to model the insecure communications. That is, instead of defining what he cannot block or spoof in a model, here, what he can do in the system is specified.
Finally, there exist a number of other channels that regulate the protocol run, such as; \emph{openElection}, \emph{closeElection}, \emph{enterBooth}, \emph{leaveBooth}, \emph{bagempty} and \emph{done}. However these will not be discussed any further.
%\newpage
\subsection{Modelling Assumptions}
\label{sec:dy:assumptions}
Although the aim in the modelling of voting systems is to obtain a model that reflects real system behaviour, there are a few assumptions that need to be made in order to avoid state explosion, which also result in abstractions in some of the features of the vVote voting system. For instance, although vVote supports the AV and STV electoral methods, FPTP will be modelled due to its simplicity in this analysis. Thus, possible privacy attacks to the system that may occur in the AV and STV electoral methods are not considered here. Additionally, in the original vVote system, ballot generation is made in a distributed fashion, which allows verifiable generation of ballot forms by distributing the trust among various entities. However, in the modelling of vVote, it is assumed that there is one honest single entity, election authority, who generates the candidate lists and digital ballot papers. This assumption can also be read as the entities that are responsible for distributed generation of ballot forms are honest and work as a single process. Similarly the WBB is a threshold-based service, which signs messages by co-operation. However, the threshold parties in the vVote model are treated as single entities.

The vVote voting system uses a mixnet to shuffle the encrypted votes cast during the election as in \pav. Previously, in~\cite{MHS14}, a CSP model of the mixnet has been given, which works as a perfect mixnet (no link between its inputs and outputs due to its non-deterministic behaviour). However, here in vVote modelling we omit this mixnet process as the WBB process already outputs the encrypted messages non-deterministically to the decryption tellers. This can also be thought as that the mixnet process is embedded in the WBB process, removing the communication between a WBB process and a mixnet one. Thus, there is no point of having two subsequent non-deterministic choices over the same inputs in terms of efficient and effective modelling. Regarding the analysis of this voting system model without a mixnet process, as the communication channels between WBB and mixnet is no blocking link because of the reasons given in Section~\ref{channels} and the messages are encrypted under authorities public key, there is not much that the intruder can do over these channels. Additionally, everything that the intruder can perform over the channel from the mixnet to the WBB can also be realised over the channel from the WBB to the decryption tellers because the messages and channel types are of the same format.

The vVote voting system employs a district information for each voter in order to allow them to vote in different constituencies. Because the modelling and analysis of this voting system does not cover this aspect of voting, the district information, used in the POD protocol, is omitted. Hence, with this abstraction, the possible privacy-related attacks to the system that may emerge with if the district information was used are not touched here.

Finally, it is assumed here that there exists only one poll worker, which opens a session for each voter with a fresh nonce. This would not impact our analysis as in the case of existence of multiple poll workers in a polling station, voters could only authenticate themselves without awaiting each other with different poll workers. However, if the cast votes were to be published on the BB one by one in the model, then there might have been issues regarding this assumption. This is because in the current model voters cast their vote in order and if the intruder could see the cast votes published on the BB in the same order, then he could violate voter anonymity.

% ---------------------------------------------------------------------------
\subsection{Honest Participants}
\label{agents}
\begin{figure}%[htp]
\centering
    \includegraphics[ clip=true, width=\textwidth, trim= 0.5cm 0cm 5cm 1cm]{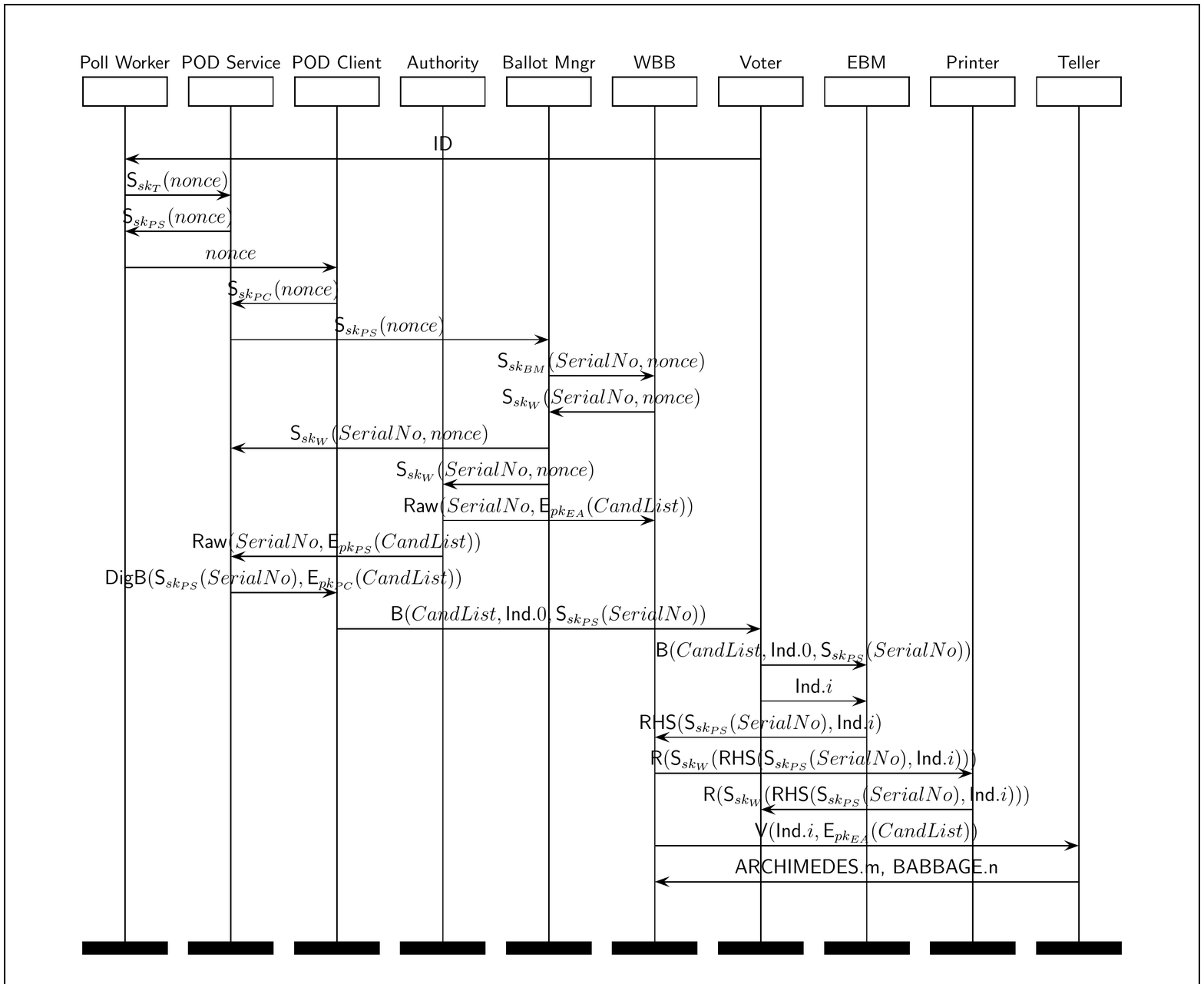} %   scale=0.40 \columnwidth
    \caption{vVote system model}
    \label{fig:systemmodel}
\end{figure}

% ---------------------------------------------------------------------------
The vVote voting system model developed for this work is defined by the processes illustrated at the top of Figure~\ref{fig:systemmodel}. All the processes are involved in the protocol by sending, receiving messages on the synchronised channels and the model behaves exactly as in Figure~\ref{fig:systemmodel}. Moreover, the model covers all phases of the vVote, including the POD protocol. 

%The following subsections present the CSP descriptions of the individual protocol participants.
%

%
% ---------------------------------------------------------------------------
\subsubsection{Voter Process} 
% ---------------------------------------------------------------------------
With the parameterised process $\it{Voter}(v,c)$, the behaviour of a voter $v \in \mathcal{V}$ voting for a chosen candidate $c \in \mathcal{C}$ is modelled. There exist two honest voters, Alice and Bob, and a misbehaving one, James, who behaves honestly in the model at first, but his secret will be shared with the intruder later on and whose communications, even the private and NSB ones, are used by the intruder.

Having authenticated herself on the NSB channel with the poll worker, Tom, the voter receives a ballot form from the POD client with the candidate list printed on it and scans her ballot data to the EBM on the secure channel, where she can see her ballot in an electronic environment. After indicating her preference by sending the index value ($\ind{i}$) to the EBM that corresponds to the candidate she wants to vote for---the index $i$ is found by using the function $\mathsf{find}(c, l)$, which finds the candidate $c$ in the sequence of candidates $l$ and defined as below, she then receives her signed receipt and leaves the polling station. 

\begin{tabbing}			
$\it{Vot}$\=$\it{er}(v,c)\defs$\\
\>\;$ openElection \then nsbcomm.v.Tom.v \then $\\
\>\;$	\underset{ \begin{subarray}{l}
                 \\ l \in \mathcal{L} 
                 \\ s \in \mathcal{S}
             \end{subarray}}{\Extchoice}
             \begin{pmatrix}\begin{array}{l} 
		scomm.podclient.v.\ballot{l}{\sign{PS}{s}}{\ind{0}} \then \\
		scomm.v.ebm.\ballot{l}{\sign{PS}{s}}{\ind{0}} \then \\
							\underset{ \begin{subarray}{l}
								 \\ i := \mathsf{find}(c, l)
             \end{subarray}}{\Extchoice} 
             \begin{pmatrix}\begin{array}{l}
    				nsbcomm.v.ebm.\ind{i} \then\\
    				nsbcomm.printer.v. \\
						\qquad\quad \receipt{W}{\sign{PS}{s}}{\ind{i}} \then\\
    				closeElection \then \STOP
              \end{array}
       \end{pmatrix}
				\end{array}
       \end{pmatrix}
 $
\end{tabbing}
where the function, \emph{find}(), is defined as 

\begin{tabbing}
$\mathsf{find}(c,l) = \begin{cases}
1 & \, \If\, c = \mathsf{head}(l)\\
1 + \mathsf{find}(c,\mathsf{tail}(l)) & \, \If\, c \neq \mathsf{head}(l)\\
\end{cases}$
\end{tabbing}

All eligible voters, Alice, Bob and James, follow this protocol, which is modelled as the parallel running of all individual voter processes synchronising on \emph{openElection} and \emph{closeElection} pairwise with the election authority. That is, each voter performs an \emph{openElection} event to begin her voting process. Each voter must also perform a \emph{closeElection} event after casting their individual vote and leaving the polling station.
\begin{tabbing}
$\emph{Voters} \defs 
{\parallel_{v,c}}\emph{Voter}(v,c)$
\end{tabbing}
%
% ---------------------------------------------------------------------------
\subsubsection{Poll Worker Process}
% ---------------------------------------------------------------------------
The poll worker, Tom, authenticates voters and starts a fresh session for each of them by choosing a nonce $n$ from the set of nonces $\mathcal{N}$. He ensures that he always authenticates a different voter, and commences a new session with a fresh nonce. The poll worker is not involved in any private communication as he only sends and receives signed nonces from and to the POD service and sends nonces to the POD client.
\flushbottom
\pagebreak
\begin{tabbing}			
$Poll$\=$worker(\mathcal{V},\mathcal{N})\defs$\\
\>$ closeElection \then \STOP$\\
\>$\extchoice$\\
\>$	\underset{ \begin{subarray}{l}
                 \\ v \in \mathcal{V}
             \end{subarray}}{\Extchoice}
\begin{pmatrix}\begin{array}{l}
 nsbcomm.v.Tom.v \then\\ 
	\underset{ \begin{subarray}{l}
                 \\ n \in \mathcal{N}
             \end{subarray}}{\Intchoice} \begin{pmatrix}\begin{array}{l} nsbcomm.Tom.podservice.\sign{T}{n}  \then 			 \\ 
  nsbcomm.podservice.Tom.\sign{PS}{n} \then 		 \\
             nsbcomm.Tom.podclient.n \then \\             
             Pollworker(\mathcal{V}\,\backprime\,\{v\}, \mathcal{N}\,\backprime\,\{n\}) % \hspace*{-1.4cm}
              \end{array}
       \end{pmatrix}
				\end{array}
       \end{pmatrix}
 $
\end{tabbing}
%
% ---------------------------------------------------------------------------
\subsubsection{Election Authority Process}
% ---------------------------------------------------------------------------
\emph{Authority} is the election authority process, which assigns a random candidate list from the list $\mathcal{L}$ for a particular serial number that has been asked for along with a nonce from the ballot manager. Following this, \emph{Authority} submits two copies of the raw ballot form: one is encrypted under the election authority's public key $pk_{EA}$ and sent to the WBB, the other is encrypted under the POD service public key $pk_{PS}$ and sent to the POD service. Hence, the WBB and the POD service keep the same candidate list associated for a particular serial number, but encrypted under different keys.
\begin{tabbing}
$\it{Authority} \qquad\quad\,\, $\=$\defs openElection \then \it{Authority_1}(\mathcal{S}, \mathcal{L})$\\

$\it{Authority_1}(\emptyset,\mathcal{L}) $\>$\defs closeElection \then \STOP$\\
			
$\it{Auth}$\=$\it{ority_1}(\mathcal{S},\mathcal{L})\defs$\\
\>$\extchoice$\\
\>$	\underset{ \begin{subarray}{l}
                 \\ s \in \mathcal{S}
                 \\ n \in \mathcal{N}
             \end{subarray}}{\Extchoice}
\begin{pmatrix}\begin{array}{l}
nsbcomm.ballotmngr.authority.\sign{BM}{s,n}  \then \\
	\underset{ \begin{subarray}{l}
                 \\ l \in \mathcal{L}
             \end{subarray}}{\Intchoice} \begin{pmatrix}\begin{array}{l} nsbcomm.authority.wbb.\raw{s}{\encrypt{EA}{l}} \\
nsbcomm.authority.podservice.\raw{s}{\encrypt{PS}{l}} \\             
             \it{Authority_1}(\mathcal{S}\,\backprime\,\{s\}, \mathcal{L}) % \hspace*{-1.4cm}
              \end{array}
       \end{pmatrix}
				\end{array}
       \end{pmatrix}
 $
\end{tabbing}
%
% ---------------------------------------------------------------------------
\subsubsection{POD Service Process}
% ---------------------------------------------------------------------------
Following a fresh session, the POD service (candidate list key sharers) receives a serial number $s$ from the ballot manager and the encrypted candidate list $\encrypt{PS}{l}$ associated with $s$ from the election authority, which is called a raw ballot. Subsequently, the digital ballot form consisting of a signed serial number and the encrypted candidate list is sent to the POD client after a transformation made on the encrypted candidate list from the POD service's public key $pk_{PS}$ to POD client's public key $pk_{PC}$.
% 
%
%\newpage
%\renewcommand{\topfraction}{0.75}
%\renewcommand{\textfraction}{0.1}
%\renewcommand{\floatpagefraction}{0.85}
\flushbottom
\pagebreak
\begin{tabbing}			
$Pod$\=$service \defs$\\
\>$ closeElection \then \STOP$\\
\>$\extchoice$\\
\>$	\underset{ \begin{subarray}{l}
                 \\ n \in \mathcal{N}
             \end{subarray}}{\Extchoice}
\begin{pmatrix}\begin{array}{l}
nsbcomm.Tom.podservice.\sign{T}{n}  \then 			 \\
nsbcomm.podservice.Tom.\sign{PS}{n}  \then 			 \\
nsbcomm.podclient.podservice.\sign{PC}{n}  \then 			 \\
nsbcomm.podservice.ballotmngr.\sign{PS}{n}  \then 			 \\
             	\underset{ \begin{subarray}{l}
                 \\ s \in \mathcal{S}
             \end{subarray}}{\Extchoice}%\hspace*{-0.5cm} 
             \begin{pmatrix}\begin{array}{l}  nsbcomm.ballotmngr.podservice.\sign{W}{s,n} \then 		 \\
             \underset{ \begin{subarray}{l}
                 \\ l \in \mathcal{L}
             \end{subarray}}{\Extchoice}
             \begin{pmatrix}\begin{array}{l}   
nsbcomm.authority.podservice.\\
\qquad\quad\raw{s}{\encrypt{PS}{l}} \\              
nsbcomm.podservice.podclient.\\
\qquad\quad \digballot{\sign{PS}{s}}{\encrypt{PC}{l}} \then \\
             Podservice
             \end{array}
       \end{pmatrix}
              \end{array}
       \end{pmatrix}
				\end{array}
       \end{pmatrix}
 $
\end{tabbing}
%
% ---------------------------------------------------------------------------
\subsubsection{POD Client Process}
% ---------------------------------------------------------------------------
The POD Client process is responsible for printing out the ballot form, which has been received as a digital ballot from the POD service (note that this should not be confused with the receipt printer). The candidate list on this digital ballot $l$ is encrypted under the POD client's public key $pk_{PC}$ with empty marking boxes denoted by $\ind{0}$. Having extracted the candidate list, the POD client prints the actual ballot form for the voter on the private channel.
\begin{tabbing}
$Pod$\=$client \defs$\\
\>$ closeElection \then \STOP$\\
\>$\extchoice$\\
\>$	\underset{ \begin{subarray}{l}
                 \\ n \in \mathcal{N}
             \end{subarray}}{\Extchoice}
\begin{pmatrix}\begin{array}{l}
nsbcomm.Tom.podclient.n \then \\ 
nsbcomm.podclient.podservice.\sign{PC}{n}  \then 			 \\
	\underset{ \begin{subarray}{l}
                 \\ s \in \mathcal{S}\\
                 \\ l \in \mathcal{L}
             \end{subarray}}{\Extchoice} \begin{pmatrix}\begin{array}{l} nsbcomm.podservice.podclient.\\ \quad \qquad\digballot{\sign{PS}{s}}{\encrypt{PC}{l}} \then \\
             	\underset{ \begin{subarray}{l}
                 \\ v \in \mathcal{V}
             \end{subarray}}{\corr{\Intchoice}}%\hspace*{-0.5cm} 
             \begin{pmatrix}\begin{array}{l}  		scomm.podclient.v.\ballot{l}{\sign{PS}{s}}{\ind{0}} \then 		 \\           
             Podclient
             \end{array}
       \end{pmatrix}
              \end{array}
       \end{pmatrix}
				\end{array}
       \end{pmatrix}
 $
\end{tabbing}
% ---------------------------------------------------------------------------
\subsubsection{Ballot Manager Process}
% ---------------------------------------------------------------------------
The ballot manager apportions the serial numbers to each ballot form uniquely and commits them to the WBB. Additionally, it also notifies the election authority and the POD service about the serial number being used.
\flushbottom
\pagebreak
\begin{tabbing}
$Ballotmanager \quad\;\,\,\, $\=$\defs openElection \then Ballotmanager(\mathcal{S})$\\
$Ballotmanager(\emptyset) $\>$ \defs closeElection \then \STOP$\\			
$Bal$\=$lotmanager(\mathcal{S})\defs$\\
\>$ closeElection \then \STOP$\\
\>$\extchoice$\\
\>$	\underset{ \begin{subarray}{l}
                 \\ n \in \mathcal{N}
             \end{subarray}}{\Extchoice}
\begin{pmatrix}\begin{array}{l}
nsbcomm.podservice.ballotmngr.\sign{PS}{n}  \then 			 \\
	\underset{ \begin{subarray}{l}
                 \\ s \in \mathcal{S}
             \end{subarray}}{\Intchoice} 
             
             \begin{pmatrix}\begin{array}{l} nsbcomm.ballotmngr.wbb.\sign{BM}{s,n}  \then 			 \\
nsbcomm.ballotmngr.authority.\sign{BM}{s,n}  \then 			 \\
nsbcomm.wbb.ballotmngr.\sign{W}{s,n}  \then 			 \\         
nsbcomm.ballotmngr.podservice.\sign{W}{s,n} \then 		 \\           
Ballotmanager(\mathcal{S} \,\backprime\, \set{s})
              \end{array}
       \end{pmatrix}
				\end{array}
       \end{pmatrix}
 $
\end{tabbing}
%
% ---------------------------------------------------------------------------
\subsubsection{The Electronic Ballot Marker (EBM)}
% ---------------------------------------------------------------------------
The EBM is a device to help voters to mark their preferences. Having received her ballot form from the POD client, the voter goes into the booth and scans her ballot form to transfer the ballot information to the EBM on the secure channel $scomm$. She then fills out the electronic ballot form on the screen by interacting with the machine and choosing the index value $\ind{i}$ corresponding to her chosen candidate. Although no one is supposed to be observing voter interaction with the EBM, it is assumed here that the index value sent from voter to the EBM can be observed by the intruder, as it will be observed anyway once she takes her receipt from the receipt printer. Afterwards, the EBM sends the marking boxes side of the ballot form, $\rhs{\sign{PS}{s}}{\ind{i}}$, to the WBB, which can then be checked by the voters.
\begin{tabbing}			
$\it{EBM} \defs $\=$ closeElection \then \STOP$\\
\>$ \extchoice $\\
\>$		
		\underset{ \begin{subarray}{l}
                 \\ l \in \mathcal{L} 
                 \\ s \in \mathcal{S}
                 \\ v \in \mathcal{V}                 
             \end{subarray}}{\Extchoice}
             \begin{pmatrix}\begin{array}{l}
		scomm.v.ebm.\ballot{l}{\sign{PS}{s}}{\ind{0}} \then \\
							\underset{ \begin{subarray}{l}
								 \\ i \in \mathcal{I}
             \end{subarray}}{\Extchoice} % \hspace*{-1.4cm}
             \begin{pmatrix}\begin{array}{l}
    				nsbcomm.v.ebm.\ind{i} \then\\
    				nsbcomm.ebm.wbb.\rhs{\sign{PS}{s}}{\ind{i}} \then \it{EBM}\\
              \end{array}
       \end{pmatrix}
				\end{array}
       \end{pmatrix}
 $
\end{tabbing}
% ---------------------------------------------------------------------------
\subsubsection{Receipt Printer Process}
% ---------------------------------------------------------------------------
The receipt printer process behaves as a typical printer, \ie it receives the receipt $r$ from the WBB, and prints it out for the voter $v$. Note that the POD client and this printer process are two different printers located in different places in the polling station.
%
%\newpage
\begin{tabbing}			
$Printer \defs $\=$ closeElection \then \STOP$\\
\>$ \extchoice $\\
\>$		
		\underset{ \begin{subarray}{l}
                  \\ v \in \mathcal{V}
                  \\ r \in receipts
             \end{subarray}}{\hspace*{-0.5cm} \Extchoice} \hspace*{-0.7cm}
             \begin{pmatrix}\begin{array}{l}
		nsbcomm.wbb.printer.r \then \\
		nsbcomm.printer.v.r \then Printer \\		
				\end{array}
       \end{pmatrix}
 $
\end{tabbing}
%
% ---------------------------------------------------------------------------
\subsubsection{The Web Bulletin Board Process}
% ---------------------------------------------------------------------------
The WBB is a public bulletin board that broadcasts the committed data during the election, such as submitted votes and signed serial numbers. Moreover, there is nothing private about this process as everything is publicly verifiable. Having received all the cast votes and sending the receipts for each voter, the WBB transfers them in the form of $\vote{\ind{i}}{\encrypt{EA}{l}}$ to the decryption teller (election key sharers) non-deterministically.

Note that a mixnet, such as re-encryption mixnet used in \pav, shuffles the cast votes arbitrarily, and outputs them in the new randomised order. In our model, this randomisation process is captured in the WBB itself, which simply outputs the votes in a non-deterministic order. Essentially this models having a perfect mixnet embedded in the WBB. This simplifies the model, and also helps in terms of reduction of the state space.

Once the teller has finished the tallying, it sends the result for each candidate (Archimedes and Babbage) to the WBB. 
\begin{tabbing}		
$\it{WBB} \qquad \quad\defs openElection \then \it{WBB}_1(\emptyset)$\\

$\it{WB}$\=$\it{B}_1(bag)\,\defs  $\\
\>$ closeElection \then \it{WBB}_2(bag)$\\
\>$\extchoice$\\
\>$	\underset{ \begin{subarray}{l}
                 \\ n \in \mathcal{N}
                  \\ s \in \mathcal{S}
             \end{subarray}}{\Extchoice}
\begin{pmatrix}\begin{array}{l}
nsbcomm.ballotmngr.wbb.\sign{BM}{s,n}  \then 			 \\
nsbcomm.wbb.ballotmngr.\sign{W}{s,n}  \then 			 \\ 
	\underset{ \begin{subarray}{l}
                 \\ l \in \mathcal{L}
             \end{subarray}}{\Extchoice} 
             
             \begin{pmatrix}\begin{array}{l}
nsbcomm.authority.wbb.\raw{s}{\encrypt{EA}{l}} \\
	\underset{ \begin{subarray}{l}
                 \\ i \in \mathcal{I}
             \end{subarray}}{\Extchoice}
     \begin{pmatrix}\begin{array}{l}
nsbcomm.ebm.wbb.\rhs{\sign{PS}{s}}{\ind{i}} \then \\
nsbcomm.wbb.printer.\\
\qquad\quad \sign{W}{\rhs{\sign{PS}{s}}{\ind{i}}} \then \\    
\it{WBB}_1(bag \union \set{\vote{\ind{i}}{\encrypt{EA}{l}}})
             \end{array}
       \end{pmatrix}
             \end{array}
       \end{pmatrix}
				\end{array}
       \end{pmatrix}
$       

\\\\$\it{WBB}_2(\emptyset) \quad\defs bage$\=$mpty \then nsbcomm.\it{Archimedes}?t_1 \then  $\\
\>$nsbcomm.\it{Babbage}?t_2 \then done \then \STOP $

\\$\it{WBB}_2(bag) \defs 	\underset{ \begin{subarray}{l}
                 \\ v \in votes
             \end{subarray}}{\Intchoice} 
             nsbcomm.wbb.teller.v \then \it{WBB}_2(bag \,\backprime\, \set{v})$
\end{tabbing}
%
% ---------------------------------------------------------------------------
\subsubsection{Decryption Teller Process}
\label{subsub:teller}
% ---------------------------------------------------------------------------
The decryption teller process---it is a thresholded setup, called decryption key sharers, but here this property is abstracted away and modelled as a single CSP process---is responsible for decrypting the votes encrypted under the election authority's public key $pk_{EA}$ and tallying them for each candidate. The results are then sent back to the WBB. What happens in the third line of the process is that because the decryption teller possesses the shared secret key $sk_{EA}$ (shared in the real system), it can decrypt and extract the candidate list of the cast votes as in $l:= \decrypt{EA}{\encrypt{EA}{l}}$. The teller then identifies for whom the vote is by checking the $i$th element of the list $l$. Accordingly, it increments the total vote received by that particular candidate by one. Once there are no more votes to tally, the teller announces the total votes for each candidate.
\flushbottom
\pagebreak
\begin{tabbing}		
$\it{Teller} \defs openElection \then \it{Teller}_1(0,0)$\\
$\it{Tel}$\=$\it{ler}_1(m,n)\defs  $\\
\>$	\underset{ \begin{subarray}{l}
                 \\ i \in \mathcal{I}
                 \\ l \in \mathcal{L}
              \end{subarray}}{\Extchoice}
\begin{pmatrix}\begin{array}{l}
nsbcomm.wbb.teller.\vote{\ind{i}}{\encrypt{EA}{l}}  \then \\             
             \begin{pmatrix}\begin{array}{l}     				
             \If nth(i,\decrypt{EA}{\encrypt{EA}{l}}) = \it{Archimedes} \Then \it{Teller}_1(m+1,n) \\
\qquad            \Else  \begin{pmatrix}\begin{array}{l} \If nth(i,\decrypt{EA}{\encrypt{EA}{l}}) = \it{Babbage} \Then \\ \qquad  \it{Teller}_1(m,n+1) \Else \STOP\\
             \end{array}
       \end{pmatrix}
             \end{array}
       \end{pmatrix}
				\end{array}
       \end{pmatrix}$\\       
\>$\extchoice$\\  
\>$ bagempty \then nsbcomm.\it{Archimedes}.m \then nsbcomm.\it{Babbage}.n \then \SKIP$
\end{tabbing}
where the function, $nth$(), is defined as 

\begin{tabbing}
$\mathsf{nth}(i,m) =
\begin{cases}
\mathsf{head}(m) & \, \If \, i = 1\\
\mathsf{nth}(i-1,\mathsf{tail}(m)) & \, \If \, i \neq 1
\end{cases}$
\end{tabbing}

\subsection{Adapting Lazy Spy}
\label{subsec:vvote:adaptation}
Lazy spy~\cite{Ros97} is an efficient intruder model as it avoids state explosion by following only its findings (deductions through the messages he has seen or from his initial knowledge). This intruder model provides active attacks against the system by not only observing the communication channels, but also blocking messages or generating and sending fake messages to any agents on the system. The framework should be altered so it can work with the cryptographic voting systems. In particular, the vVote voting system model is equipped with a number of voting system specific messages as well as the cryptographic ones (see Figure~\ref{fig:messages}). Hence, the existence of these messages requires further deduction rules that need to be defined so that the intruder can act as he is supposed to regarding those messages. Secondly, the initial knowledge of the intruder $\mathcal{IK}$ is also model specific, hence, it needs to be defined according to the voting system model and as this set of knowledge is used to specify what the intruder knows and what he can learn, it needs to be defined carefully. Lastly, because of the introduction of various channel types in the analysis of voting systems, the intruder model needs to be amended so that the private channels stay private and NSB channels are, indeed, not blocked or spoofed by the intruder.

In order to allow the intruder to compose messages, there are a number of deduction rules. Recall that a deduction is a pair $(X,f)$, where $X$ is a finite set of facts and $f$ is the fact that can be generated, providing that the intruder possesses $X$ and these inferences are denoted as $X \hence f$. It should be ensured that the intruder deals with a finite set of facts because FDR cannot handle an infinite number of states. Thus, arbitrary nesting of encryptions and sequences needs to be avoided. To do so, the set of data-types are limited to the types that are enough to build protocol messages. Although the intruder can generate ``bad'' facts (objects that are not of the form real messages sent among protocol agents), these facts will do him no good~\cite{RSG+00}. That is because the agents in the protocol can only communicate with the messages that they understand---the messages need to be in the same form as they are expected. Hence, the deduction rules with which the intruder is able to build and decompose all protocol messages are adequate for the analysis.

In Section~\ref{sec:analysis}, we shall need to introduce an algebraic equivalence over messages in order to model anonymity correctly, and ensure that the intruder cannot draw conclusions about voters' actions based on the contents of encryptions for which he does not have the decryption key. Equivalences are not considered in~\cite{RSG+00}, and for good reason: they are liable to cause trouble with the argument used there to justify a strong typing assumption. However, it should be noted that this equivalence we shall introduce will not be used in the deduction rules that the intruder can use to generate new knowledge; it is used only in the specification. It cannot, therefore, be used by the intruder to break the typing system.

The deduction rules regarding this analysis $\mathcal{D}$ are the union of deductions defined in Table~\ref{tab:deducemvvote}. The original framework provides the deduction rules regarding cryptographic primitives, and the rest is specific to the vVote voting system. The new deduction rule BALLOT-COMP enables ballot forms to be composed if the intruder possesses the set $\set{l,\sign{}{s}, \ind{i}}$, where $l$ is the candidate list, $s$ is serial number and $\ind{i}$ is the index value, corresponding to the chosen candidate and conversely the deduction rule BALLOT-DCMP helps the intruder to decompose ballot forms and obtain all the data on it. Similarly, the intruder can also work on any composition and decomposition of any other messages in the model. For instance, RHS-COMP and RHS-DCMP are the deduction rules related to cast ballot forms, consisting of an index value and a signed serial number $\set{\ind{i}, \sign{}{s}}$. VOTE-COMP and VOTE-DCMP are the deduction rules related to the votes, in the form of $\vote{\ind{i}}{\encrypt{}{l}}$ (note that these do not contain a serial number). The deduction rules regarding the digital ballots, consisting of a signed serial number and an encrypted candidate list, $\set{\sign{}{s}, \encrypt{}{l}}$, are DIG.BLT-COMP and DIG.BLT-DCMP. Similarly, RAW.BLT-COMP and RAW.BLT-DCMP are the two deduction rules that help the intruder compose and decompose the raw ballots,$\set{s, \encrypt{}{l}}$, and IND-COMP and IND-DCMP are the index related deduction rules. Hence, with this set of deduction rules, $\mathcal{D}$, the intruder is enabled to deduce messages that are used to attack the protocol. 

\begin{table}%[htp]
\small
\centering
\begin{tabular}{|llll|}
 \hline
 & & & \\
SYM-ENC. & $ \set{k, m}  $ & $\hence $ & $\sencrypt{}{m}$ \\ 
SYM-DEC. & $ \set{k, \sencrypt{}{m}}  $ & $\hence $ & $m$ \\
 & & & \\ 
ASYM-ENC. & $ \set{pk, m}  $ & $\hence $ & $\encrypt{}{m}$ \\ 
ASYM-DEC. & $ \set{sk, \encrypt{}{m}}  $ & $\hence $ & $m$ \\
 & & & \\ 
SIGN-SIG. & $ \set{sk, m}  $ & $\hence $ & $\sign{}{m}$\\
SIGN-EXT. & $ \set{pk, \sign{}{m}}  $ & $\hence $ & $m$\\
 & & & \\
BALLOT-COMP. & $ \set{l,\sign{}{s}, \ind{i}}  $ & $\hence $ & $ \ballot{l}{\sign{}{s}}{\ind{i}} $ \\ 
BALLOT-DCMP. & $ \set{\ballot{l}{\sign{}{s}}{\ind{i}}}  $ & $\hence $ & $l , \sign{}{s} , \ind{i}$\\
 %& & & \\
 & & & \\ 
RHS-COMP. & $ \set{ \ind{i}, \sign{}{s}}  $ & $\hence $ & $ \rhs{\ind{i}}{\sign{}{s}} $ \\  
RHS-DCMP. & $ \set{\rhs{\sign{}{s}}{\ind{i}}}  $ & $\hence $ & $  \ind{i} , \sign{}{s}$\\
 & & & \\ 
VOTE-COMP. & $ \set{\ind{i}, \encrypt{}{l}}  $ & $\hence $ & $ \vote{\ind{i}}{\encrypt{}{l}} $ \\
VOTE-DCMP. & $ \set{\vote{\ind{i}}{\encrypt{}{l}}}  $ & $\hence $ & $ \ind{i} , \encrypt{}{l}$\\
 & & & \\ 
DIG.BLT-COMP. & $ \set{\sign{}{s}, \encrypt{}{l}}  $ & $\hence $ & $ \digballot{\sign{}{s}}{\encrypt{}{l}}$ \\ 
DIG.BLT-DCMP. & $ \set{\digballot{\sign{}{s}}{\encrypt{}{l}}}  $ & $\hence $ & $\sign{}{s} , \encrypt{}{l}$\\
 & & & \\ 
RAW.BLT-COMP. & $ \set{s, \encrypt{}{l}}  $ & $\hence $ & $\raw{s}{\encrypt{}{l}}$ \\ 
RAW.BLT-DCMP. & $ \set{\raw{s}{\encrypt{}{l}}}  $ & $\hence $ & $s , \encrypt{}{l}$\\
 & & & \\ 
IND-COMP. & $ \set{i}  $ & $\hence $ & $ \ind{i}$ \\ 
IND-DCMP. & $ \set{\ind{i}}  $ & $\hence $ & $ i$\\
 & & & \\ 
 \hline
\end{tabular}
\caption{Deduction rules capturing the properties of cryptographic primitives and the vVote voting system messages.}
\label{tab:deducemvvote}
\end{table}

As mentioned earlier, the set $comms$ needs to be defined for all messages in the model illustrated in Figure~\ref{fig:messages} so that the intruder can justify that a message being heard is actually from a real communication between agents. As in the protocol, no agent sends any message to himself, for such communications are ensured to be omitted with $a \neq b$ below, which also implies that if an agent sends a message to himself, it cannot be blocked or spoofed by the intruder.
\begin{tabbing}
$\it{comms} = \{a.b.m \mid m \leftarrow \mathcal{M}, a\leftarrow \mathcal{A}, b\leftarrow \mathcal{A}, a \neq b\}$
\end{tabbing}
The messages in the model that make sense to the intruder are: $comms$, all the messages from real communications, $\it{Nsbcomms}$, the set of messages that cannot be blocked or spoofed by the intruder, and $\it{Ucomms}$, the set of insecure messages that the intruder can act as in the Dolev-Yao intruder model~\cite{DY83}.

As all honest participants communicate on the NSB channels, not including a message type in the set $\it{Nsbcomms}$ means that the intruder cannot even overhear that kind of message. Hence, the messages in the form of a ballot are not included in this set, as the intruder should not be able to observe any communication involving a ballot form between honest participants (denoted as \emph{commBallots}). For example, a voter scanning her ballot form to the EBM should not be observed by the intruder and this is how he is prevented from overhearing and blocking the private channels.
\begin{tabbing}
$Nsbcomms     = comms \,\backprime\, commBallots$~\footnotemark
\end{tabbing}
\footnotetext{We denote the set subtraction as $\,\backprime\,$ in order to distinguish it from the hiding operator ``$\backslash$'' in CSP.}
The insecure messages that the intruder can overhear, block or use in any way in the line of Dolev-Yao model, are defined with the set $\it{Ucomms}$ as follows. It should be noted that the set in the analysis of vVote covers all the messages that are communicated by the dishonest voter James.
%
%\newpage
\begin{tabbing}
$\it{Ucomms} = \Union($\=$\{q.q'.f \mid q.q'.f \leftarrow comms, q \leftarrow  \set{\emph{James}}, q'\leftarrow  agents \},$\\ 
\>$ \{q.q'.f \mid q.q'.f \leftarrow  comms, q \leftarrow  agents, q'\leftarrow  \set{\emph{James}} \})$
\end{tabbing}

The set \emph{Ucomms} can be extended with any set of information. For instance, the insecure communications in \emph{Ucomms} do not yet include the receipts taken by the voters during the election. Hence, the intruder cannot cannot block the voters' taking their receipts as they are still on the no blocking channel. However, if we add the set of receipts that can be taken away by any voters (denoted as \emph{commReceipts}) to the set \emph{Ucomms}, then the intruder could also block the voters taking their receipts because in such case receipt information would flow on the insecure communication channels. However, it should be noted that the more information is given to the intruder, the longer the automated verification takes due to the increased number of deductions made by the intruder.

The adaptation of the intruder model to voting systems analysis has been made by introducing different channel types, introduced in this section, and the CSP definition of the lazy spy intruder model is kept intact. Further details about the lazy spy intruder model, which is called $Intruder$ here and defined in terms of the channels $learn$ and $say$, can be found in~\cite{Ros97}.

\subsection{Putting the Network Together}\label{subsec:puttingnetwork} 
\begin{figure}%[htp]
\centering
    \includegraphics[scale=0.55]{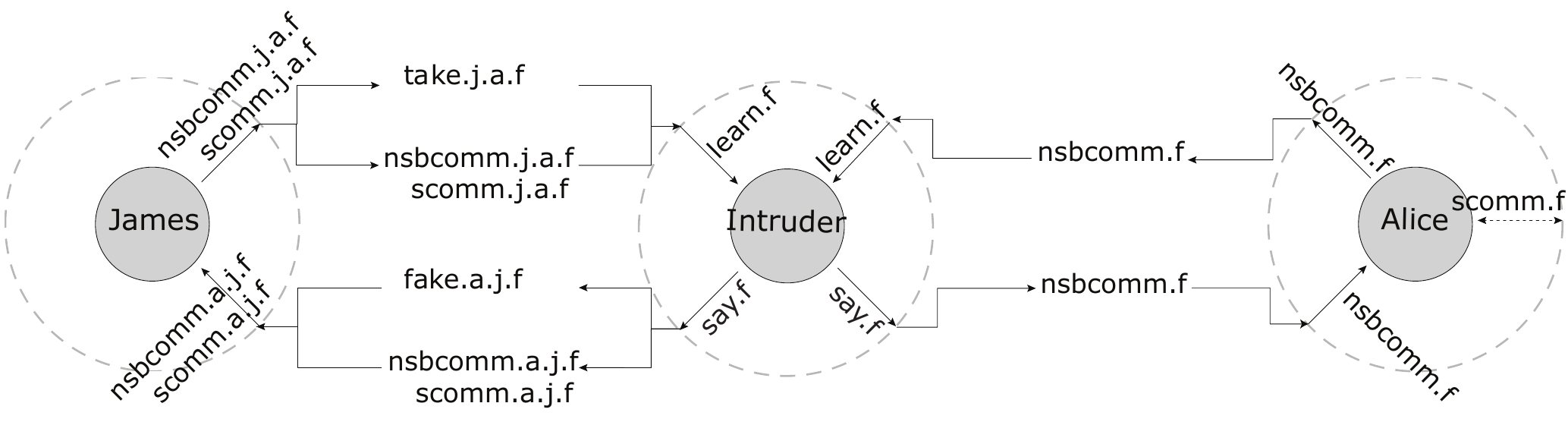} %[width=\columnwidth] or [scale=0.88]
    \caption{The lazy spy intruder model}
    \label{fig:channels}
\end{figure}

Figure~\ref{fig:channels} illustrates how the intruder is connected to the dishonest voter James, and the honest voter Alice, whereby Alice's private channel $scomm$ is kept private, but her insecure NSB channels can be observed by the intruder, whereas all the channels of James are under the control of the intruder. That is, the intruder can overhear all insecure communications acting as a medium, but he can only intercept and fake the messages in the form of insecure data \emph{Ucomms} (it defines all communications from and to James) as defined in the previous section. Moreover, he has no power over the private channels of the honest voters. 

The processes that construct the voting system model and the intruder model are connected by using the renaming operator. That is, $nsbcomm.a.b.m$ and $learn$ channels are renamed to a $take$ channel, and the $nsbcomm.b.a.m$ and $say$ channels are renamed to a $\it{fake}$ channel from the agent $a$'s point of view. Similarly, the intruder process is also renamed and the aim is to connect the intruder with the agents. Hence, the intruder channel $learn.m$ is mapped to the events of the form $take.a.b.m$, and $say.m$ is renamed to $\it{fake}.a.b.m$. To this end, a renaming function for the process $P$ and agent name $p$ can be defined as follows:
\begin{tabbing}
$r(P,p) \defs P$\=$ \rensubs{nsbcomm.p, take.p}{nsbcomm.p, nsbcomm.p} $\\
\>$ \rensubs{nsbcomm.a.p, fake.a.p}{nsbcomm.a.p, nsbcomm.a.p \mid a \in \mathcal{A}} $\\
\>$ \rensubs{scomm.p, take.p}{scomm.p, scomm.p} $\\
\>$ \rensubs{scomm.a.p, fake.a.p}{scomm.a.p, scomm.a.p \mid a \in \mathcal{A}} $
\end{tabbing}
Hence, the renamed voter process for the voter $v$, for instance, can be defined as follows. Note that, the private channel $scomm$ is renamed to the $take$ and $fake$ channels, because this models the malicious behaviour of a corrupt voter.
\begin{tabbing}
$r(\it{Vot}$\=$er(v,c),v) \defs $\\
\>$ \it{Vot}$\=$er(v,c) $\\
\>\>$ \rensubs{nsbcomm.v, take.v}{nsbcomm.v, nsbcomm.v} $\\
\>\>$ \rensubs{nsbcomm.a.v, fake.a.v}{nsbcomm.a.v, nsbcomm.a.v \mid a \in \mathcal{A}} $\\
\>\>$ \rensubs{scomm.v, take.v}{scomm.v, scomm.v} $\\
\>\>$ \rensubs{scomm.a.v, fake.a.v}{scomm.a.v, scomm.a.v \mid a \in \mathcal{A}} $
\end{tabbing}
Similarly, the other processes that construct the vVote voting system model are renamed as in the above example. Consequently, the voting system model, $\it{Model}$, which is ready to be modified by the intruder, is defined as the parallel composition of all those renamed processes.
\begin{tabbing}
$\it{Model} \defs $\=$
\it{rVoters}
\parallel
rPollworker
\parallel 
rAuthority
 \parallel
rEBM
\parallel
rPodservice $\\
\>$\parallel 
rPrinter 
\parallel
rBallotmanager 
\parallel
rPodclient 
\parallel
rWBB
\parallel
rTeller$
\end{tabbing}
The parallel composition (interface parallel) above is constructed in a way that the processes only synchronise on the $nsbcomm$ and $scomm$ channels on which they send messages to each other, leaving the insecure channels ($take$, $fake$) vulnerable to be used by the intruder. The following shows how two processes are put in an interface parallel and therefore, the above parallel composition of $\it{Model}$ should be constructed in this way.
\begin{tabbing}
$\it{Mod}$\=$el \defs \it{rVoters}
\underset{ X }{\parallel}
rPollworker
\parallel 
\dots$
\end{tabbing}
\begin{tabbing}
where $X =  \eset{
$\=$nsbcomm.v.Tom,  nsbcomm.Tom.v,$\\ 
\>$scomm.v.Tom, scomm.Tom.v \,\mid\, v \leftarrow \mkset{V}
}$
\end{tabbing}
Similarly, the intruder process is prepared by renaming as below so that the intruder can overhear the messages on the insecure NSB channels (\emph{Nsbcomms}) and act as the Dolev-Yao intruder on the insecure channels (\emph{Ucomms}). For the vVote analysis, these sets are defined in the previous section. 
%
%\newpage
\begin{tabbing}
$rInt$\=$ruder \defs $\\
\>$ Intr$\=$uder $\\
\>\>$\rensubs{say, learn}{say, say}$ \\
\>\>$ \rensubs{nsbcomm.p.p'.f, take.q.q'.f}{learn.f, learn.f \mid 
\begin{array}{l}
p.p'.f \in \it{Nsbcomms},\\
q.q'.f \in  \it{Ucomms},\\
p\neq p', q \neq q' 
\end{array}}$\\
%p.p'.f \in Nsbcomms, q.q'.f \in  Ucomms, p\neq p', q \neq q' }$\\
\>\>$ \rensubs{fake.p.p'.f}{say.f \mid p.p'.f \in \it{Ucomms},\, p\neq p'}$
\end{tabbing}
The process $System_{\it{vVote}}$ is then defined in terms of the parallel composition of \emph{Model} and \emph{rIntruder}, which synchronise on the channels they share. 
\begin{tabbing}
$System_{\it{vVote}} \defs \it{Model} \underset{  \left\{ \begin{array}{|c|}
 nsbcomm, take, fake
\end{array} \right\} }\parallel \it{rIntruder}$
\end{tabbing}
Having modelled the system, it is ready to be automatically analysed under the anonymity requirement in the next section.

% ---------------------------------------------------------------------------
\section{vVote Analysis}
\label{sec:analysis}
% ---------------------------------------------------------------------------
%% ---------------------------------------------------------------------------
%\section{Automated Anonymity Verification}
 %\label{sec:dy:analysis}
%% ---------------------------------------------------------------------------
In this section, the first fully-automated analysis of the vVote voting system is presented under a Dolev-Yao intruder model and using the weak anonymity definition given in~\cite{MHS14} as the specification. It requires that when the two channels $c.x$ and $d.x$ are swapped over for all values of $x$, if the resulting process is indistinguishable from the original one, then the process provides anonymity. In the previous section, the system where the honest agents of the voting system model and the intruder interact has been modelled as the process $System_{\it{vVote}}$. In this system, the process \emph{rVoters} is defined as the parallel composition of the \emph{rVoter}() processes for each of the voters: Alice, Bob and James. However, the anonymity specification in this analysis is approached in a different way whereby the two systems, which are expected to be indistinguishable, are defined as two separate system behaviours without using the renaming operator. Hence, the following processes, namely, $\it{rVoters_1}$ and $\it{rVoters_2}$ model the two different voters' behaviour, which may result in two different system behaviours: on the one hand Alice votes for Archimedes, Bob votes for Babbage and James can vote for either Archimedes or Babbage, whereas on the other hand, Alice votes for Babbage, Bob votes for Archimedes and again James can vote for any of them. The resulting two different system behaviours are $\it{{System'}_{vVote}}$ and $\it{{System''}_{vVote}}$, respectively. Note that the misbehaving voter, James, shares his knowledge with the intruder, thus his behaviour is modelled with the renamed voter process \emph{rVoter}(), which allows the intruder to use his knowledge, whereas the honest voters Alice and Bob are modelled using the honest voter model \emph{Voter}(). However, the NSB channels can still be observed by the intruder, meaning that the intruder acts passively on these channels.
\begin{tabbing}
$\it{rVoters_1} \defs $\=$(\it{Voter}(Alice,Archimedes) \interleave \it{Voter}(Bob, Babbage))$\\ \>$\interleave (\it{rVoter}(James, Archimedes) \intchoice \it{rVoter}(James, Babbage))$\\
\\
$\it{rVoters_2} \defs $\=$(\it{Voter}(Alice,Babbage) \interleave \it{Voter}(Bob, Archimedes))$\\ \>$\interleave (\it{rVoter}(James, Archimedes) \intchoice \it{rVoter}(James, Babbage))$
\end{tabbing}
The systems are modelled in such a way that the intruder can see everything James does, including his private messages, whilst Bob and Alice can vote freely without any interception/blocking or spoofing. That is, although the intruder can still overhear the public channels and inference from those messages, Alice and Bob vote under the private and NSB channel assumptions.

The observational equivalence that is used for the analysis necessitates masking of the encrypted values in order to avoid false positive attacks, as the intruder can distinguish two ciphertexts, even if he does not know the secret key---this was not a case in the NSPK~\cite{NS78} analysis as the secrecy specification is different from the anonymity used in this analysis. To this end, a masking function $\mathsf{maskFact}$ is deployed. The function renames all messages encrypted under a public key, whose corresponding secret key is not known by the voter, to a data $ciphertext$ and if the secret key is in the intruder's initial knowledge, then he is allowed to differentiate two ciphertexts by not masking them.
\begin{tabbing}
$\mathsf{mask}$\=$\mathsf{Fact}(\encrypt{}{m}) = \If\, \mathsf{dual}(pk) \in \mathcal{IK} \Then
                                    \encrypt{}{m} \Else ciphertext $  
\end{tabbing}
The masking function $\mathsf{mask}(P)$ can also be defined for the processes, which masks all encrypted facts of a given process, $P$, using the $\mathsf{maskFact}$ function for all the data that appears in this process. (No keys are ever sent over the network, so if the intruder does not know a secret key at the beginning, he will not learn it later.)
\begin{tabbing}
$m$\=$ask(P) \defs $\\
\>$P $\=$ \rensubs{chnl.a.a'.\digballot{\sign{}{s}}{\mathsf{maskFact}(\encrypt{}{l}})}{chnl.a.a'.\digballot{\sign{}{s}}{\encrypt{}{l}}}$\\
\>\>$\rensubs{chnl.a.a'.\raw{s}{\mathsf{maskFact}(\encrypt{}{l}})}{chnl.a.a'.\raw{s}{\encrypt{}{l}}}$\\
\>\>$\rensubs{chnl.a.a'.\vote{\ind{i}}{\mathsf{maskFact}(\encrypt{}{l}})}{chnl.a.a'.\vote{\ind{i}}{\encrypt{}{l}}}$
\end{tabbing}
where $chnl \in \set{nsbcomm, take, fake}$, the serial number $s \in \mathcal{S}$, the candidate list $l \in \mathcal{L}$ and the index value $i \in \mathcal{I}$.

After applying the masking function to both $\it{{System'}_{vVote}}$ and $\it{{System''}_{vVote}}$, they are ready for the analysis under the anonymity specification. To this end, the anonymity requirement of this voting system model is checked with the following trace equivalence in which the private channels are hidden.
\begin{tabbing} 
$\mathsf{mask}(\it{{System'}_{vVote}}) \hide \eset{scomm} \tequiv \mathsf{mask}(\it{{System''}_{vVote}}) \hide \eset{scomm}$
\end{tabbing}
FDR verifies that the two systems refine each other, meaning that they are trace equivalent and hence that the intruder cannot distinguish them. As a result, the vVote voting system model provides anonymity under the Dolev-Yao intruder model. 

% --------------------------------------------------------------------------- 
 \subsection{Analysis under Alternative Assumptions}
 \label{sec:dy:alternative}
% ---------------------------------------------------------------------------
Although, the framework used in the previous section provides a firm comprehensive foundation for analysis of voting systems, it is also important to see whether the framework supports further extensions to those assumptions made previously, because one of the important challenges in electronic voting systems may be to maintain requirements even under the assumption of the corrupt agents, for instance, misbehaving participants. Such analyses are possible with slight modifications to the voting system and the intruder models. The following paragraphs present two of these analyses of the vVote under different assumptions.

\paragraph{\bf{Corrupt POD Service}} The POD service is an important part of the print-on-demand protocol. It receives raw ballot data including a serial number $s$ and candidate list encrypted under $pk_{PS}$, and sends the digital ballot by signing the serial number to the POD client. If the POD service is corrupt, which is modelled as that the POD service's secret is possessed by the intruder, the raw ballot received by the POD service, say $\raw{s_3}{\encrypt{PS}{\myseq{\it{Archimedes}, \it{Babbage}}}}$, can be captured and decrypted by the intruder. Hence, the intruder can extract the candidate list $\myseq{\it{Archimedes}, \it{Babbage}}$ and deduce its association with the serial number $s_3$. Following this, when he observes that Alice's receipt with the index value $\ind{1}$ has the serial number $s_3$ on it, he is then able to infer that Alice has voted for the first candidate of the candidate list $\myseq{\it{Archimedes}, \it{Babbage}}$, which is Archimedes. Therefore, the intruder distinguishes the two systems as Alice cannot have voted for Babbage. This counter-example is produced by FDR automatically and illustrated by the following partial trace.
%
%\newpage
\begin{tabbing}
$\mytrace{\dots$\\
$nsbcomm.authority.wbb.\raw{s_3}{ciphertext},$\\
$nsbcomm.authority.podservice.\raw{s_3}{\encrypt{PS}{\myseq{\it{Archimedes}, \it{Babbage}}}},$\\
$nsbcomm.podservice.podclient.\digballot{\sign{PS}{s_3}}{ciphertext},$\\
$enterBooth.Alice,$\\
$nsbcomm.Alice.ebm.\ind{1}}$
\end{tabbing}
It can be observed from the above trace that the intruder cannot decrypt the ciphertext in the message sent from the authority to the WBB, as it is encrypted under the authority's public key $pk_{EA}$, which is seen as \emph{ciphertext} in the trace.

This scenario emphasises the importance of the single point failure in the protocol security. In the real system, however, the POD service is thresholded, meaning that all threshold parties, sign, encrypt or decrypt messages jointly, without any party learning the ballot order. Therefore, the above would be a threat against vVote, should all threshold parties collude.

\paragraph{\bf{Corrupt Authority}} A similar approach can be taken to model a corrupt election authority, who leaks sensitive information that can be used by the intruder. Since the authority is responsible for assigning random candidate lists to each requested serial number from the ballot manager, the candidate list encrypted under the authority's public key will be revealed when he is corrupt. Therefore, the intruder's accurate deduction about the candidate lists would violate voter anonymity by revealing the candidate list of a ballot form used by a particular voter. The following trace produced by FDR demonstrates that when the authority is compromised, which is modelled as the intruder knows his secret key $sk_{EA}$, the intruder violates Alice's anonymity by deducing how she has voted. In more detail, the intruder can overhear the candidate order $\myseq{\it{Archimedes}, \it{Babbage}}$ on Alice's ballot form before she casts her vote. Once Alice indicates her preference by the index value $\ind{1}$, the chosen candidate, Archimedes, is revealed to the intruder. 
\begin{tabbing}
$\mytrace{\dots$\\
$enterBooth.Alice$\\
$nsbcomm.authority.wbb.\raw{s_1}{\encrypt{PS}{\myseq{Archimedes, Babbage}}},$\\
$nsbcomm.authority.podservice.\raw{s_1}{ciphertext},$\\
$nsbcomm.podservice.podclient.\digballot{\sign{PS}{s_1}}{ciphertext},$\\
$nsbcomm.Alice.ebm.\ind{1}}$
\end{tabbing}
Although no one is supposed to be observing voter interaction with the EBM, it is assumed here that the index value $\ind{i}$ sent from voter to the EBM can be observed, as it will be observed anyway once she takes her receipt from the receipt printer. Thus, the two counter-examples above were found by FDR when the intruder could observe these index values. If the intruder was not allowed to do so, the counter-examples would still appear once the voter has taken her receipt in a protocol run. Moreover, the two counter-example traces above include only $nsbcomm$ events, which illustrates that the intruder does not need to block or spoof messages on those channels and hence a passive observer possessing the corresponding secret keys would also be able to attack the system.

There are numerous corruption scenarios one can think of and that can be modelled and analysed using this framework. In particular, the two presented here emphasise the importance of the case of a corrupt single entity, such as the election authority and POD service, where the voters are at a high risk of losing their anonymity. The vVote voting system has a solution to these problems, to some extent, by having the ballot forms generated by the threshold election authorities. However, if the other trusted entities, like the EBM, are acting dishonestly, the system is vulnerable to various attacks. Additionally, it was observed that a corrupt WBB does not reveal anything useful for the intruder to break the anonymity requirement of the system, because the WBB is public anyway.

%The next section investigates the modelling and analysis of the secrecy requirement for voting systems.
% --------------------------------------------------------------------------- 
 \subsection{Secrecy Analysis using Lazy Spy}	
 \label{sec:dy:secrecy}
% ---------------------------------------------------------------------------
The lazy spy intruder model~\cite{RG97} was used to verify the authentication and secrecy requirements of security protocols. In this analysis, a secret is defined as the terms $\set{AtoB, BtoA}$, and the intruder model is defined so that when the intruder learns one of the secrets in a protocol run, the process flags it up using the channel $intruderknows$. When this event occurs in a protocol run, the secret is not secret any more. Previously, in the anonymity requirement analysis of vVote, this event was omitted because such an event was not needed for the formal specification of this requirement. However, perhaps not for the paper-based voting systems, where the voters are generally not required to use public key pairs to encrypt their votes, but especially in remote voting systems, this specification can be used to verify whether the voting systems maintain the secrecy of the votes. That is, it can be verified whether the intruder ever gets to know a secret originated by a particular voter or any other agent. Moreover, the secret data can be defined more specifically for each voting system, such as, a candidate encrypted and cast by the voter as in the FOO scheme, in which a voter encrypts her vote, blinds the encrypted version and sends it to the registrar. To this end, the highlighted expression below is added to the intruder model to flag up the intruder's knowledge about a secret $f$ from the set of secrets \emph{Banned}. 
\begin{tabbing}
$\it{Ignorantof(f)} \defs $\=$  f \in \mathcal{M}\, \&\, learn.f \then \it{Knows}(f)$\\
						\>$  \extchoice \it{infer}?t \in \{(X,f') \mid\, $\=$(X,f') \in \mathcal{D}, f' = f\} \then \it{Knows}(f) $
\end{tabbing}
\begin{tabbing}
$\it{Knows}(f) \defs $\=$ f \in \mathcal{M} \,\&\,  say.f \then \it{Knows}(f)$ \\
	\>$\extchoice f \in \mathcal{M} \,\&\,  learn.f \then \it{Knows}(f)$ \\
	\>$\extchoice \it{infer}?t \in \{(X,f') \mid\, $\=$(X,f') \in \mathcal{D}, f \in X \} \then \it{Knows}(f)$\\
	\>$\extchoice \bim{f} \in \bim{Banned}\, \& \, \bim{intruderknows.f} \then \bim{Knows(f)}$
\end{tabbing}
Consequently, for the secrecy specification of a voting system, the following trace refinement needs to be checked.
\begin{tabbing}
$\STOP \trefinedby System \hide \Sigma \,\backprime\, \eset{intruderknows}$
\end{tabbing}
where $\Sigma = \eset{nsbcomm, take, fake, intruderknows}$, the alphabet of the process $System$.
%
% ---------------------------------------------------------------------------
\section{Discussion and Conclusion}
\label{sec:conclusion}
% ---------------------------------------------------------------------------
%% ---------------------------------------------------------------------------
 %\section{Discussion}
 %\label{sec:dy:conclusion}
%% ---------------------------------------------------------------------------
In the beginning of modelling the intruder capabilities for voting systems, a need for different channel types was mentioned. Regarding this, the need came out when the model was initially analysed under the full Dolev-Yao intruder model that can overhear, intercept and spoof any messages on all channels other than the private channels. From this initial analysis, the following counter-example was produced, which shows that with such an intruder the vVote voting system is open to anonymity attacks, which verifies the observation made in~\cite{KR05} about the FOO voting system.
\begin{tabbing}
$\mytrace{\dots$\\
$\mathbf{scomm.podclient.Alice.\ballot{\myseq{Archimedes,Babbage}}{\sign{PS}{s_1}}{\ind{0}},}$\\
$comm.Alice.ebm.\ind{1},$\\
$\mathbf{scomm.podclient.Bob.\ballot{\myseq{Archimedes,Babbage}}{\sign{PS}{s_2}}{\ind{0}},}$\\
$comm.Bob.ebm. \ind{2},$\\
$closeElection,$\\
$comm.wbb.teller.\vote{\ind{2}}{ciphertext},$\\
$take.wbb.teller.\vote{\ind{1}}{ciphertext},$\\
$comm.teller.wbb.Archimedes.0}$
\end{tabbing}
What the intruder does in the counter example trace above is to block or intercept with the channel $take$ all the other votes except Bob's. In this case, Alice has voted for Archimedes with the index value $\ind{1}$ and Bob has voted for Babbage with $\ind{2}$ on the private channels---the candidate orders on the private channels $scomm$ are hidden in the analysis and they are revealed here just for illustration. Once the election is closed, tallying starts and the votes are transferred from the WBB to the teller, the intruder intercepts the vote with the index value $\ind{1}$, and waits until Bob's vote is counted. Having seen that no one has voted for Archimedes, the intruder then deduces that Bob has voted for Babbage. This is a genuine and generic attack---not only to vVote, but it is applicable to any voting system. However, as it is not possible in a real system that the intruder can block all votes but one, it was assumed in the analysis that at least two honest votes are tallied at the end of the election. On the other hand, the intruder works fully on James' messages on the public and private channels as if he votes in public.

Having modified the system with the adaptation of insecure NSB channels, it was verified that the vVote voting system provides anonymity. This, together with the corrupt agent scenarios, demonstrated that the abstract models and formal definitions of requirements are adequate for the automatic verification of voting system protocols. Additionally, it was shown that the active intruder model modelled in this article is much more powerful in terms of mounting various kinds of attacks than the passive attacker model used in the previous analyses~\cite{MHS14, MHS15}, which can only observe the messages on the public channels.

Table~\ref{tab:vvotevertimes} illustrates the verification times of the automated analysis of vVote voting system based on the efficient models. In the table, the restricted Dolev-Yao (D-Y) is the intruder model that is restricted to only a subset (James's communications) of all messages, whereby he can act as in the Dolev-Yao intruder model. The restriction is modelled with the existence of private and NSB channels. Additionally, the full D-Y model is where the intruder can act maliciously on all channels, but the private ones---voters' privacy is still maintained. However, the refinement does not hold, which necessitates the NSB channels in the model. The restricted and full D-Y results cannot be compared with each other, as the verification times vary depending on the voters' being honest or dishonest, they give some idea about how large a model FDR can handle before state explosion for each test. 
%
%\begin{table}
%\small
%\centering
%\begin{tabular}{||c|c|r|r||c|c|r|r||}
%\hline
	%\multicolumn{4}{||c||}{\bf{Restricted D-Y}} & \multicolumn{4}{c||}{\bf{Full D-Y}}\\\hline\hline
  %&  \multicolumn{1}{c|}{Refine} & \multicolumn{1}{c|}{States} & \multicolumn{1}{c||}{Time} &  & \multicolumn{1}{c|}{Refine} & \multicolumn{1}{c|}{States} & \multicolumn{1}{c||}{Time} \\\hline
%3v 2c  & $ \tick $ &$16,063,214$ &$20m 29s$ & 2v 2c & X & $1,040,462$ & $1m 15s$ \\\hline
%4v 2c  & $-$ &$-$ &$-$ & 3v 2c & $-$ & $-$ & $-$ \\\hline
%\end{tabular}
%\caption{The FDR verification times for vVote (``$-$'' means no result is produced in a reasonable time)}
%\label{tab:vertimesvvote}
%\end{table}
%%\rowcolor{gray!25} \cellcolor{gray!25}
\setlength{\extrarowheight}{.75ex}
\begin{table}%[htp] %{}
\centering
\begin{tabular}{c|ccc | c|ccc}
\multicolumn{4}{c}{{Restricted D-Y}} & \multicolumn{4}{c}{{Full D-Y}}\\ \hline
 &  \multicolumn{1}{r}{Refine} & \multicolumn{1}{r}{States} & \multicolumn{1}{r}{Time} &  & \multicolumn{1}{c}{Refine} & \multicolumn{1}{c}{States} & \multicolumn{1}{c}{Time} \\ \hline
3v 2c  & $ \tick $ &$16,063,214$ &$1h 14m 56s$ & 2v 2c & X & $899,494$ & $1m 45s$ \\ 
 3v 3c  & $-$ &$-$ &$-$ & 2v 3c & X & $5,040,658$ & $22m 26s$ \\
 4v 2c  & $-$ &$-$ &$-$ & 2v 4c & $-$ & $-$ & $-$ \\
 4v 3c  & $-$ &$-$ &$-$ & 3v 2c & $-$ & $-$ & $-$ \\
\end{tabular}
\caption{The FDR verification times for vVote. As the required state space grows quickly with the number of voters and candidates, it was not possible to produce results in some cases as FDR cannot handle with such huge states. Those are denoted as ``$-$'' in the table.}
\label{tab:vvotevertimes}
\end{table} % colortbl is used 

In this article, we have proposed an efficient and flexible formal approach to modelling and analysis of cryptographic voting systems. In order to validate the suitability of the framework, the vVote voting system was analysed against an anonymity specification. To do so, an extensive number of other such rules regarding voting systems have been defined. These enable the intruder to learn and deduce further from his knowledge so as to able to use it to break the protocol objectives. Moreover, we introduced special channel types, private and NSB channels, in order to reason about voting systems under appropriate assumptions, as it has been observed that no voting system model can provide anonymity under an unrestricted Dolev-Yao intruder model. The framework can be applied to other voting systems providing that a CSP model of the system that is compatible with the framework is produced, and system-specific deduction rules are given. %Intuitively, the results should scale up with the number of agents, but more thoughts need to be made. %The automated verifications were conducted using the FDR model checker, and it was shown that FDR together with Roscoe's lazy spy is adequate in the analysis of such systems thanks to the latter's efficient structure, thus avoiding unnecessary inferences. 
\subsection{Acknowledgements} Some of the work was conducted while both authors were at the University of Surrey and carried out under the EPSRC-funded trustworthy voting systems (TVS) project. Additionally, the authors would like to thank Steve Schneider, Chris Culnane and David M. Williams for their useful discussions on the technical content.

\section*{Acronyms}
\label{sec:acronyms}

\begin{tabbing}
AKISS \hspace*{1.9cm} \= Active Knowledge in Security Protocols.\\
AV \> alternative vote.\\
BM \> ballot manager.\\
CA \> Certificate Authority \\
CSP \> Communicating Sequential Processes.\\
DRE \> Direct Recording by Electronics.\\
EA \> election authority.\\
E2E \> end-to-end.\\
EBM \> electronic ballot marker.\\
FDR \> Failures-Divergence Refinement.\\
FPTP \> first-past-the-post.\\
InS \> insecure channel. \\
IRV \> instant-runoff voting.\\
NB \> no blocking.\\
NOH \> no overhearing. \\
NS \> no spoofing. \\
NSB \> no spoofing and blocking.\\
NSPK \> Needham-Schroeder Public-Key.\\
POD \> print-on-demand.\\
PC \> print-on-demand client. \\
PS \> print-on-demand service. \\
RHS \> right hand side.\\
SBA \> short ballot assumption.\\
STV \> single transferable vote.\\
VEC \> Victorian Electoral Commission.\\
WBB \> web bulletin board.
\end{tabbing}
% Acknowledgements for colleagues, referees, ...



%% Numbered

%% Numbered without titles
%\bibliographystyle{model1a-num-names}

%% Harvard
%\bibliographystyle{model2-names} \biboptions{authoryear}

%% Vancouver numbered
%\usepackage{numcompress}
%\bibliographystyle{model3-num-names}

%% Vancouver name/year
%\usepackage{numcompress}\bibliographystyle{model4-names}\biboptions{authoryear}

%% APA style
% \bibliographystyle{model5-names}\biboptions{authoryear}

%% AMA style
%\bibliographystyle{model6-num-names}

%% `Elsevier LaTeX' style
%\bibliographystyle{elsarticle-num}
\bibliographystyle{alpha}
\bibliography{References}
%\newpage
%\input{7_Appendices}
\end{document}